\documentclass[article,aps,prx,twocolumn,nofootinbib,longbibliography]{revtex4-1}
\usepackage{graphicx}
\usepackage{amssymb}
\usepackage{amsmath}
\usepackage{bm}
\usepackage{color}
\usepackage[utf8]{inputenc}
\usepackage{chngcntr}
\usepackage{hyperref}
\usepackage[caption=false]{subfig}

\usepackage{float}
\usepackage{physics}
\usepackage{amsfonts}

\usepackage{amssymb}
\usepackage{braket}

\usepackage{bm}
\usepackage{xcolor}

\begin{document}

\title{Morphology of three-body quantum states from machine learning}

\author{David Huber}
\affiliation{Technische Universit\"{a}t Darmstadt, Department of Physics, Institut f\"{u}r Kernphysik, 64289 Darmstadt, Germany}

\author{Oleksandr V. Marchukov}
\affiliation{Technische Universit\"{a}t Darmstadt, Institut  f\"{u}r Angewandte Physik, Hochschulstra{\ss}e 4a, 64289 Darmstadt, Germany}

\author{Hans-Werner Hammer}
\affiliation{Technische Universit\"{a}t Darmstadt, Department of Physics, Institut f\"{u}r Kernphysik, 64289 Darmstadt, Germany}
\affiliation{ExtreMe Matter Institute EMMI and Helmholtz Forschungsakademie Hessen für FAIR (HFHF), GSI Helmholtzzentrum f\"{u}r Schwerionenforschung GmbH,
64291 Darmstadt, Germany}
\author{Artem G. Volosniev}
\affiliation{Institute of Science and Technology Austria, Am Campus 1, 3400 Klosterneuburg, Austria}

\begin{abstract}
    The relative motion of three impenetrable particles on a ring,
    in our case two identical fermions and one impurity,
    is isomorphic to a triangular
    quantum billiard. Depending on the ratio $\kappa$ of the impurity and
    fermion masses, the billiards can be integrable or non-integrable (also referred to in the main text as chaotic). To set the stage, we first
    investigate the energy level distributions of the billiards as a
    function of $1/\kappa\in [0,1]$
    and find no evidence of integrable cases beyond the limiting values
    $1/\kappa=1$ and $1/\kappa=0$. Then, we use machine learning tools
    to analyze properties of probability distributions of individual
    quantum states.
    We find that convolutional neural networks can correctly classify
    integrable and non-integrable states.
    The decisive features of the wave functions are the normalization and
    a large number of zero elements, corresponding to the existence of a
    nodal line. The network achieves typical accuracies of 97\%,
    suggesting that machine learning tools can
    be used to analyze and classify the morphology of probability densities obtained in
    theory or experiment.
\end{abstract}

\maketitle

\section{Introduction}
The correspondence principle conjectures that highly excited states of a quantum system carry information about the classical limit~\cite{Born1957}. In particular, it implies that there must be means to tell a difference between a `typical' high-energy quantum state that corresponds to an integrable classical system from a `typical' high-energy state that corresponds to a chaotic system. A discovery of such means is a complicated task that requires a coherent effort of physicists, mathematicians, and philosophers~\cite{Berry1987,Belot1997,Zurek2003}. Currently, there are two main approaches to study chaotic features in quantum mechanics. One approach relies on the statistical analysis of the energy levels of a quantum-mechanical system. Another focuses on the morphology of wave functions.  These approaches led to a few celebrated conjectures that postulate features of energy spectra and properties of eigenstates~\cite{Berry1977a,Berry1977b, Bohigas1984}. The postulates are widely accepted now, thanks to numerical as well as experimental data~\cite{Stockmann1999, Haake2001}.   

Numerical and experimental data sets produced to confirm the proposed conjectures are so large that it is difficult, if not hopeless, for the human eye to find universal patterns beyond what has been conjectured. 
Therefore, it is logical to look for computational tools that can learn (with or without supervision) universal patterns from large datasets.  
One such tool is deep learning (DL)~\cite{LeCun2015},  
  which is a machine learning method that uses artificial neural networks with multiple layers for a progressive learning of features from the input data.
  It requires very little engineering by hand, and can easily be used to analyze big data across disciplines, in particular in physics~\cite{carleo2019}.
   DL tools present an opportunity to go beyond the
  standard approaches of quantum chaologists~\cite{Berry_1989}. For example, in this paper, neural networks built upon many states are used to analyze the morphology of individual wave functions. Therefore, DL provides us with means to connect and extend tools already used to understand `chaos' in quantum mechanics.

Recent work~\cite{Kharkov2020} has already opened an exciting possibility to study the quantum-classical correspondence in integrable and chaotic systems using DL. In particular, it has been suggested that a neural network (NN) can learn the difference between wave functions that correspond to integrable and chaotic systems. It is important to pursue this research direction further, and to understand  and interpret how a NN distinguishes the two situations. This information can be used in the future to formulate new conjectures on the role of classical chaos in quantum mechanics. The main challenge here is the extraction of this information from a NN, which often resembles a black box. Ongoing research on interpretability of NNs suggests certain routes to understand the black box~\cite{Guidotti2018, olah2018the, Zhang2018} (see also recent works that discuss this question for applications in physical sciences~\cite{Kaspschak:2020ezh,kaspschak2020neural,Dawid2020}). 
However, there is no standard approach to this problem. In part, this is connected to the fact that DL relies on general-purpose learning procedures, therefore, one does not expect 
that there can be a unique way to analyze a neural network at hand. For example, as we will see, the training of a network for the `integrable' vs `chaotic' state recognition is very similar to the classic dog-or-cat classification.\footnote{The dog-or-cat classifier in this case is a network with one output label for a dog and one for a cat, which has been trained using a set of a few thousand pictures.} It is not clear, however, that the tools that can be used to interpret the latter (e.g., based on stable spatial relationships~\cite{Zhang2020}) are also useful for the former. In particular, a training set for the `integrable'-or-`chaotic' problem contains information about vastly different length scales (determined by the energy), whereas a training set for cats vs dogs has only length scales given by the size of the animal. Therefore, it is imperative to study interpretability of neural networks used in physics separately from that in  other applications.

In this paper we analyze a neural network, which has been trained using highly excited states of a triangular billiard, and attempt to extract the learned features. Billiards are conceptually simple systems, yet it is expected that they contain all necessary ingredients for studying the role of chaos in quantum mechanics~\cite{Stockmann1999}. Furthermore, eigenstates of quantum
billiards are equivalent to the eigenstates of the Helmholtz equation with the corresponding boundary conditions, which connects quantum billiards and the wave chaos in microwave resonators~\cite{Richter1999,Stockmann1999}.
The triangular billiard is one of the most-studied models in quantum chaology~\cite{Berry1984_triangle, Li1985, Kaufman1999,Aquiar2008, Lima2013}, and therefore it is well-suited for our study focused on analyzing neural networks as a possible tool for quantum chaology. 

In our analysis, we rely on convolutional neutral networks (ConvNets) for image classification~\cite{Rawat2017}, which have recently been successfully applied to categorize numerical and experimental data in physical sciences~\cite{Wang2016,Broecker2017, Hu2017,zhang2019, rem2019, bohrdt2019, Pekalski2020, Kharkov2020,Rz_dkowski_2020}.
These advances motivate us to apply ConvNets to categorize quantum states as integrable and non-integrable. Our goal can be stated as follows: given a set of highly excited states, build a network that can classify any input state as integrable or not, and, moreover, study features of this network. One comment is in order here. There are various definitions of quantum integrability~\cite{Caux_2011}, so we need to be more specific. In this work, we call a quantum system integrable, if it is Bethe-ansatz integrable, i.e., if one can write any eigenstate as a finite superposition of plane waves. We shall also sometimes use the word chaotic instead of non-integrable. Finally, we note that the properties
  `integrable' and `non-integrable' are usually attached to a given
  physical system, e.g., following an analysis of global properties like
  the distribution of energy levels. However, the correspondence principle
  implies that these labels can also be applied to individual states
  of a quantum system. In this paper, we use both notions and show that
  they are compatible. We employ neural networks to analyze the wave
  functions of individual quantum states.

We show that a trained network accurately classifies a state as being `integrable' or `non-integrable', which implies that a ConvNet learns certain universal features of highly-excited states. We argue that a trained neural network considers almost any random state generated by a Gaussian, Laplace or other distribution as `chaotic', as long as the state includes a sufficient amount of zero values.
This observation agrees with our intuition that a non-integrable state has only weak correlations.
We discuss the effect of the noise and coarse graining in our classification scheme, which sets limitations on the applicability of neural network to analyze experimental and numerical data.    

Our motivation for this work is thus threefold:
First, we want to demonstrate that neural networks
can classify the morphology of the three-body states correctly.
Therefore, we investigate a known model system with two identical
fermions and an impurity as a function of the impurity mass in order
to be able to verify the neural network analysis.
Second, we want to analyse the classifying network and understand
the way it operates.
Our third goal is to use the network analysis to clarify the situation
regarding suggested new integrable cases for other parameter values
than the established ones~\cite{Li1995, McGuire2001}. 
However, we do not find any evidence of such cases.

The paper is organized as follows. In Sec.~\ref{sec:system} we introduce the system at hand: a triangular quantum billiard that is isomorphic
  to three impenetrable particles on a ring. Its properties are discussed
  in Sec.~\ref{sec:properties} using standard methods. 
  In Sec.~\ref{sec:network}, we present our neural network approach
  and use it in Sec.~\ref{sec:results} to classify the states of the system.
  Moreover, we analyze the properties of the network.
  In Sec.~\ref{sec:concl}, we conclude. Some technical details are presented
  in the appendix.

\section{Formulation}
\label{sec:system}

We study billiards isomorphic to the relative motion of three impenetrable particles in a ring: two fermions and one impurity. Characteristics of these triangular billiards are presented below, see also Ref.~\cite{Krishnamurthy1982, glashow1997}. Our choice provides us with a simple parametrization of triangles in terms of the mass ratio, $\kappa=m_I/m$, where $m_I$ ($m$) is the mass of the impurity (fermions). 
Furthermore, it allows us to shed light on the problem of three particles in a ring
with broken integrability~\cite{Lamacraft2013,Barfknecht2015,joseph2016}.

For simplicity, we always assume that the impurity is heavier than (or as heavy as) the fermions, corresponding to $1/\kappa\in [0,1]$.
As we show below, this leads to a family of isosceles triangles with the limiting cases $(90^\circ,45^\circ,45^\circ)$ for $1/\kappa=0$ and $(60^\circ,60^\circ,60^\circ)$ for $\kappa=1$. 
These limiting triangles correspond to two identical hard-core particles in a square well and a 2+1 Gaudin-Yang model on a ring~\cite{Guan2013}, respectively. Both limits are Bethe-ansatz integrable, see Refs.~\cite{Li1985,Schachner1994} for a more detailed discussion. 
Note that certain extensions to the Bethe ansatz suggest that additional solvable cases exist~\cite{Li1995, McGuire2001}. However, our numerical analysis does not 
find any traces of solvability beyond the two limiting cases, and supports the widely accepted idea that 
almost any one-dimensional problem with mass imbalance is non-integrable (notable exceptions are discussed in Refs.~\cite{Olshanii2015,Loft2015, Scoquart2016,Harshman2017, Liu2019}). 
Therefore, in this work we refer to systems with $1/\kappa=0$ and $1$ as integrable, in the sense that they can be analytically solved using the Bethe ansatz (cf.~Ref.~\cite{Caux_2011}). Systems with other mass ratios are called non-integrable.

\subsection{Hamiltonian}

\begin{figure}
\includegraphics[scale=0.9]{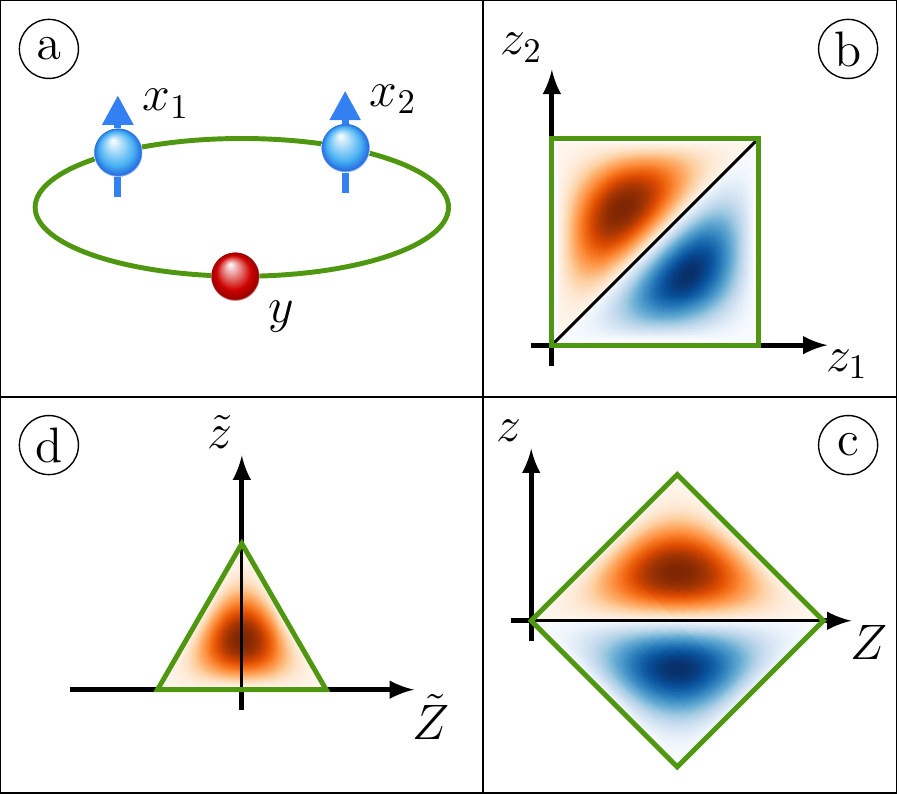}
\caption{The figure illustrates the system of interest and the correspondence between three particles in a ring and a triangular billiard.
Panel (a): Three particles in a ring. Two fermions have coordinates $x_1$ and $x_2$, the coordinate of the impurity is $y$, see Eq.~(\ref{eq:ham}). 
Panel (b): The coordinates $z_1$ and $z_2$ describe the relative motion of three particles. Panel (c): The coordinates $z$ and $Z$, which are obtained after rotation of $z_1$ and $z_2$, see Sec.~\ref{sec:system} B).
Panel (d): The triangular billiard is obtained upon rescaling the coordinates $z$ and $Z$. To illustrate the transformation from (a) to (d), we sketch the (real-valued) ground-state wave function
for $\kappa=1$ in panels (b)-(d). The blue (red) color denotes negative (positive) values of $\psi_{P=0}$ (see.~Eq.~(\ref{eq:phi_p})), which is chosen to be real in our analysis. 
The intensity matches the absolute value. Note that in panel (d), we show only $\tilde z>0$. The wave function for $\tilde z<0$ is obtained from the fermionic symmetry by reflection on the $\tilde Z$ axis.}
\label{fig:figure_1}
\end{figure}

The Hamiltonian of a three-particle system with zero-range interactions reads as
\begin{equation} 
H=-\frac{\hbar^2}{2m}\frac{\partial^2}{\partial x_1^2}-\frac{\hbar^2}{2m}\frac{\partial^2}{\partial x_2^2}-\frac{\hbar^2}{2\kappa m}\frac{\partial^2}{\partial y^2} +g\sum_i \delta(x_i-y).
\label{eq:ham}
\end{equation}
Everywhere below we focus on the limit $g\to \infty$. In Eq.~(\ref{eq:ham}), $0<x_i<L$ ($0<y<L$) is the coordinate of the $i$th fermion (impurity), while $L$ is the length of the ring, see Fig.~\ref{fig:figure_1} (a). 
The eigenstates ($\phi$) of $H$ are periodic functions in each variable. They are antisymmetric with respect to the exchange of fermions, i.e.,
$\phi(x_1,x_2,y)=-\phi(x_2,x_1,y)$.  Furthermore,
the limit $g\to \infty$ demands that $\phi$ vanishes when the fermion approaches the impurity, i.e., $\phi(x_i\to y)\to 0$. For convenience,  we use the system of units in which $\hbar=1$ and $m=1$ in the following. For our numerical analysis, we choose units such that $L=\pi$.

The Hamiltonian $H$ can be written as a sum of the relative and center-of-mass parts.  
To show this, we expand $\phi$ using a basis of non-interacting states, i.e., 
\begin{equation}
\phi(x_1,x_2,y)=\sum_{n_1,n_2,n_3} a_{n_1,n_2}^{(n_3)}e^{-\frac{2\pi i}{L}(n_1 x_1 + n_2 x_2 + n_3 y)},
\end{equation}
where $a_{n_1,n_2}^{(n_3)}=-a_{n_2,n_1}^{(n_3)}$ to satisfy antisymmetric condition on the wave function.
It is straightforward to see that the Hamiltonian does not couple states with different values 
  of the ``total momentum'', $P=2\pi\frac{n_{\mathrm{tot}}}{L}; n_{\mathrm{tot}}=n_1+n_2+n_3$ because of translational invariance. For example, the operator $\delta(x_1-y)$ couples two states via the matrix element:
\begin{equation}
\int \mathrm{d}x_1\mathrm{d}x_2\mathrm{d}y \delta(x_1-y)e^{-\frac{2\pi i}{L}(n_1-n_1') x_1 + (n_2-n_2') x_2 + (n_3-n_3') y)},
\end{equation}
which equals $\delta_{n_2,n_2'}\delta_{n_1+n_3,n_1'+n_3'}$, and, hence, conserves~$P$.
The integral of motion, $P$, allows us to write the wave function as 
\begin{equation}
\phi=e^{-i P y} \sum_{n_1,n_2}a_{n_1,n_2}^{(n_{\mathrm{tot}}-n_1-n_2)}e^{-\frac{2\pi i}{L}(n_1 (x_1-y) + n_2 (x_2-y) )},
\end{equation}
and define the function, which depends only on the relative coordinates:
\begin{equation}
  \label{eq:phi_p}
\psi_P(z_1,z_2)=e^{i P y}\phi(x_1,x_2,y),
\end{equation} 
where $z_i=L\theta(y-x_i)+x_i-y$, with the Heaviside step function: $\theta(x>0)=1, \theta(x<0)=0$.
The coordinates $z_i$ are chosen such that the function $\psi_P(z_1,z_2)$ takes values on $z_i \in[0,L]$, 
see Fig.~\ref{fig:figure_1}~(b).

The function $\psi_P$ is an eigenstate of the Hamiltonian
\begin{equation}
H_P=-\frac{1}{2}\sum_{i=1}^2 \frac{\partial^2}{\partial z_i^2}-\frac{1}{2\kappa}\left(\sum_{i=1}^2 \frac{\partial}{\partial z_i}\right)^2
+i\frac{P}{\kappa}\sum_{i=1}^2 \frac{\partial}{\partial z_i},
\label{eq:h_p}
\end{equation}   
which will be the cornerstone of our analysis. As we show below, it is enough to consider only $H_{P=0}$ for our purposes. To diagonalize $H_0$, we resort to exact diagonalization in a suitable basis. As a basis element, we use the real functions $\sin \left(\frac{n_1 \pi z_1}{L}\right)\sin \left(\frac{n_2 \pi z_2}{L}\right) - \sin \left(\frac{n_1 \pi z_2}{L}\right)\sin \left(\frac{n_2 \pi z_1}{L}\right)$, where $n_1$ and $n_2$ are integers with $n_{max}>n_1>n_2>0$, which is a standard choice for this type of problems, see, e.g., \cite{Miltenburg1994,Dehkharghani2016}. Our choice of the basis ensures that $\psi_{P=0}$ is real for the ground and all excited states. The parameter $n_{max}$ defines the maximum element beyond which the basis is truncated. Note that the basis element is the eigenstate of the system for $1/\kappa=0$. Therefore, we expect exact diagonalization to perform best for large values of $\kappa$ and more poorly for $\kappa=1$. To estimate the accuracy of our results, we benchmark against the exact solution for an equilateral triangle ($\kappa=1$), see the discussion in the Appendix. Using $n_{max}=130$, we calculate about 4000 states whose energies have relative accuracy of the order of $10^{-3}$. This set of 4000 states is an input for our analysis in the next section. 

To summarize this subsection: We perform the transformation from $H,\phi$ to $H_P,\psi_P$ to eliminate the coordinate of the impurity from the consideration. 
Our procedure can be considered as the Lee-Low-Pines transformation~\cite{Lee1953} in coordinate space, which is a known tool for studying many-body systems with impurities in a ring~\cite{Volosniev2017, Panochko2019, Mistakidis2019, Jager2019}. Below we argue that $H_P$ can be further mapped onto a triangular billiard. Note however that we are going to work with $H_P$ everywhere. Its eigenfunctions are defined on a square (see Fig.~\ref{fig:figure_1}~(b)), allowing us to use them directly as an input for ConvNets.

\subsection{Mapping onto a triangular billiard}

It is known that three particles in a ring can be mapped onto a triangular billiard~\cite{Krishnamurthy1982, glashow1997}. 
Here we show this mapping starting with $H_P$. First of all we rotate the system of coordinates to eliminate the mixed derivative $\frac{\partial}{\partial z_1}\frac{\partial}{\partial z_2}$; see Fig.~\ref{fig:figure_1} (c). 
To this end, we introduce the system of coordinates $z=(z_2-z_1)/\sqrt{2}$ and $Z=(z_2+z_1)/\sqrt{2}$, where the Hamiltonian reads as
\begin{equation}
  \label{eq:H_P}
H_P(z,Z)=-\frac{1}{2}\frac{\partial^2}{\partial z^2}-\frac{1}{2}\frac{\partial^2}{\partial Z^2}-\frac{1}{\kappa}\frac{\partial^2}{\partial Z^2}
+i\frac{\sqrt{2}P}{\kappa}\frac{\partial}{\partial Z}.
\end{equation}   
The last term here can be eliminated by a gauge transformation $\psi_P\to \exp\left(\frac{i\sqrt{2}P}{\kappa+2}Z\right)\psi_P$. 
Therefore, in what follows we only consider $P=0$ without loss of generality.  
We shall omit the subscript, i.e., we write $\psi$. Note that it is enough
to study only $z\geq 0$, because of the symmetry of the problem. 

To derive the standard 
Hamiltonian for quantum billiards:
\begin{equation}
h=-\frac{1}{2}\frac{\partial^2}{\partial \tilde z^2}-\frac{1}{2}\frac{\partial^2}{\partial \tilde Z^2},
\end{equation} 
we rescale and shift the coordinates as $\tilde z =z$, and $\tilde Z=\sqrt{\kappa/(\kappa+2)}(Z-L/\sqrt{2})$, see Fig.~\ref{fig:figure_1} (d). The Hamiltonian $h$ is defined on an isosceles triangle with the base angle obtained from 
$\tan(\alpha)=\sqrt{(\kappa+2)/\kappa}$. For systems with more particles the corresponding transformations $H\to H_P\to h$ lead to quantum billiards in polytopes, allowing one to connect an $N$-body quantum mechanical problem to a quantum billiard in $N-1$ dimensions. This can be a route for finding new applications of solvable models, see Refs.~\cite{Krishnamurthy1982,Olshanii2015,Scoquart2016}. 

Finally, we note that if the interaction term in Eq.~(\ref{eq:ham}) was an impenetrable wall with some radius $R$ instead of the delta function, then the considerations above would also lead to a mapping of the system onto a triangle. (See~Ref.~\cite{Hill1980} for an illustration with an equilateral triangle.)

\section{Properties of the System}
\label{sec:properties}
A discussion of highly excited states of triangular billiards can be found in the literature~\cite{Berry1984_triangle, Li1985, Kaufman1999,Aquiar2008, Lima2013}. 
However, we find it necessary to review some known results and calculate some new quantities in order to explain our current understanding of the difference between integrable and non-integrable states. In principle, highly excited states of a quantum system can be simulated using microwave resonators (see, e.g.,~\cite{Stein1990,Heller1992}), or generated by means of Floquet engineering -- by choosing the driving frequency to match the energy difference between the initial and the desired final state (see, e.g., Ref.~\cite{Lenz_2011}). Therefore the results of this section are not of purely theoretical interest, as they can be observed in a laboratory.

As we outlined in the introduction, there are two main approaches for analyzing a connection between highly-excited states and classical integrability. The first one relies on statistical properties of the energy spectra, while the second one focuses on the morphology of individual quantum states. This section sets the stage for our further study by discussing these approaches in more detail.

\subsection{Energy} 
\begin{figure}
\vspace{-1.1cm}
\includegraphics[scale=0.55]{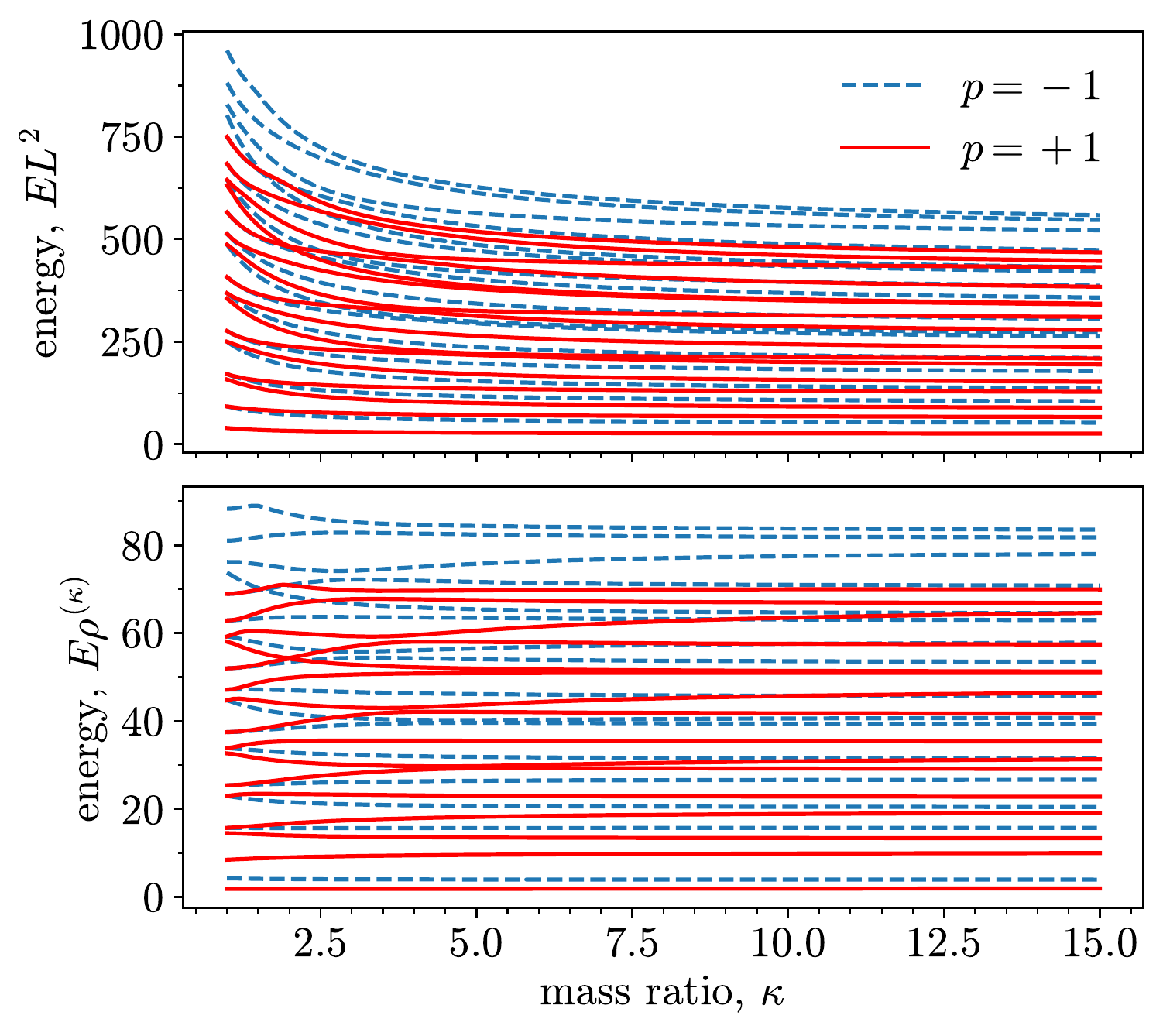}
\caption{The spectrum of the Hamiltonian $H_{P}$
    from Eq.~(\ref{eq:H_P}) as a function of the mass ratio, $\kappa$. The figure presents the first 30 states: 15 mirror-symmetric states, $p=1$, (red, solid curves), and 15  mirror-antisymmetric states, $p=-1$, (blue, dashed curves). The top panel shows the eigenstates of Eq.~(\ref{eq:h_p}). The bottom panel shows these energies multipled by $\rho^{(\kappa)}$ from Eq.~(\ref{eq:rho_kappa}).}
\label{fig:figure_2}
\end{figure}

We start by calculating the energy spectrum. It provides a basic understanding of the evolution from an `integrable' to a `chaotic' system in our work as a function of $\kappa$.
We present the first 30 states of $H_0$ in Fig.~\ref{fig:figure_2} (top). Note that an isosceles triangle has a symmetry axis ($\tilde Z \to -\tilde Z$),
which corresponds to a mirror transformation (in the particle picture this symmetry corresponds to $z_{i}\to L-z_{i}$). 
 The wave function can be symmetric or antisymmetric with respect to the mirror transformation and we consider these cases separately.  The former states are denoted as having $p=1$,
and the latter have $p=-1$.

The energy spectrum features inflation of the spacing between levels, which can be understood as a repulsion of levels in the Pechukas gas~\cite{Pechukas1983,Yukawa1985}. According to Weyl's law, it can also be  interpreted as a change of the density of states, $\rho^{(\kappa)}(E)=\mathrm{d}N/\mathrm{d}E$, where $N(E)$ is the number of states with the energy less than $E$. The function $\rho^{(\kappa)}(E)$ can be easily calculated using Weyl's law~\cite{Ivrii2016} for the triangular billiard described by the Hamiltonian $h$:
\begin{equation}
\rho^{(\kappa)}(E\to \infty)\to \frac{L^2}{4\pi}\sqrt{\frac{\kappa}{\kappa+2}}.
\label{eq:rho_kappa}
\end{equation}  
The density of states is independent of the energy in this equation because we work with a two-dimensional object. Equation~(\ref{eq:rho_kappa}) is derived assuming large values of $E$, however, in practice, it also describes well the density of states in a lower part of spectrum (cf.~Refs.~\cite{Kaufman1999}). 
If we multiply the energies presented in Fig.~\ref{fig:figure_2} (top) by $\rho^{(\kappa)}$ then we obtain a spectrum without inflation, i.e., all levels  are equally spaced on average, see Fig.~\ref{fig:figure_2} (bottom). 

Multiplication of $E$ by $\rho^{(\kappa)}$ is a simple example of unfolding, which allows us to directly compare features of the energy spectrum for different values of $\kappa$. 
The goal of the unfolding is to extract the ``average'' properties of the levels distribution and, thus, diminish the effect of local level density fluctuations in the spectrum. While there are many possible ways to implement the unfolding procedure, which depend on the properties of the energy spectrum (for further information see, e.g., Refs.~\cite{bohigas1984a, prosen1993, Haake2001}), the ultimate goal is to obtain rescaled levels with unit mean spacing. Below, we rescale all of the energy levels by the mean distance between them, thus, obtaining the unit mean spacing. We benchmarked results of this unfolding against more complicated approaches, and found qualitatively equivalent outcomes.

\begin{figure}
\includegraphics[scale=0.55]{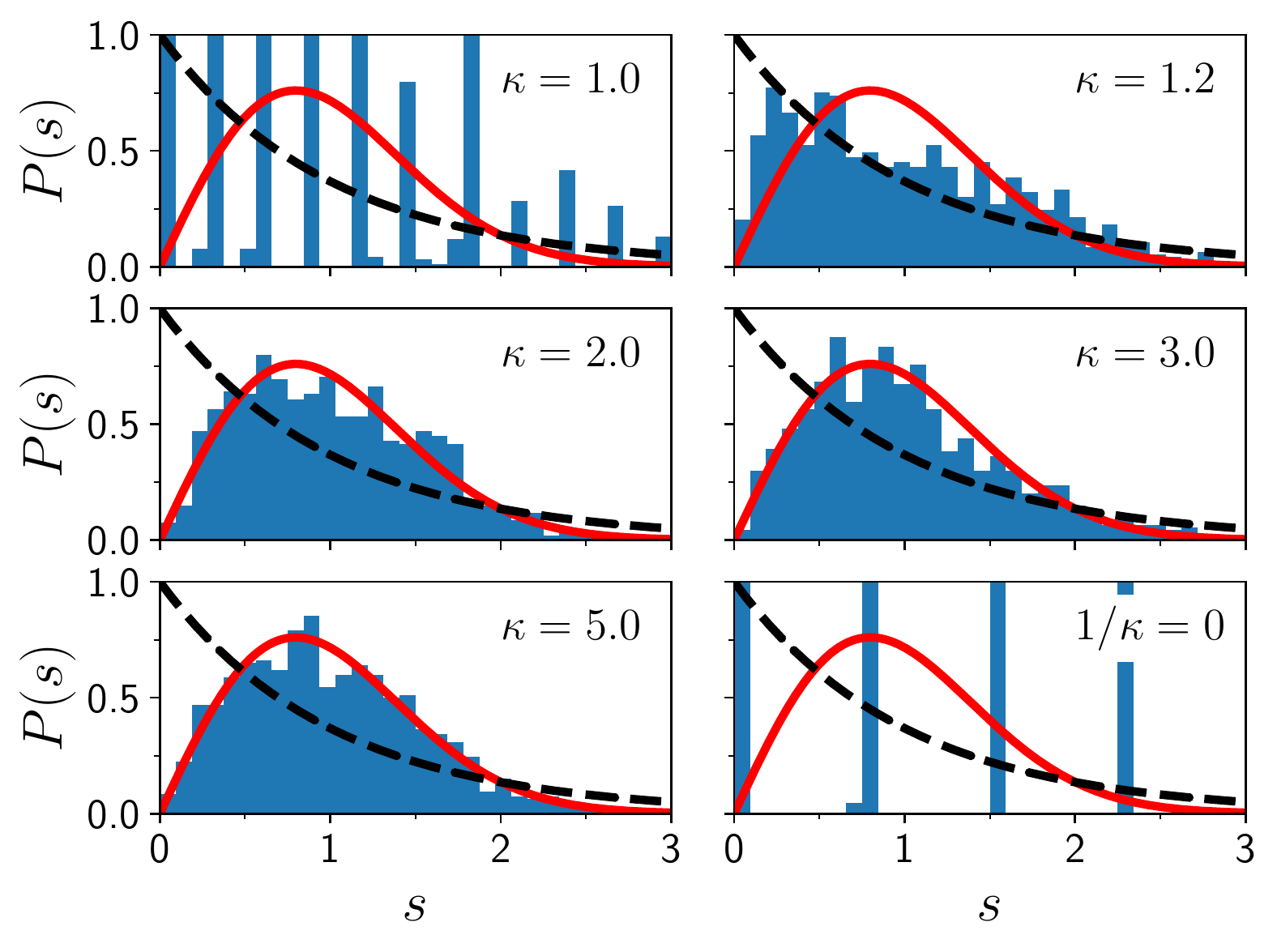}
\caption{The histogram shows the nearest-neighbor distribution, $P(s)$, as a function of $s$ for different values of the mass ratio, $\kappa$. The (red) solid curve shows the Wigner distribution from Eq.~(\ref{eq:p_goe}). The (black) dashed curve shows the Poisson distribution, $e^{-s}$. To produce these figures, only states with $p=1$ are used.}
\label{fig:figure_3}
\end{figure}

We use unfolded spectra to analyze the distribution of nearest neighbors, $P(s)$, which shows the probability that the distances between a random pair of two neighboring energy levels is $s$. The function $P(s)$ is presented in~Fig.~\ref{fig:figure_3}, see also~\cite{Schachner1994, Kaufman1999,Aquiar2008}, where some limiting cases are analyzed. For the sake of discussion, we only study the states with $p=1$, however, we have checked that the case with $p=-1$ leads to qualitatively identical results. 
The size of bins in the histograms in~Fig.~\ref{fig:figure_3} is virtually arbitrary~\cite{prosen1993}. For visual convenience, we have followed a rule of thumb that the number of bins should be taken at approximately a square root of the number of the considered levels. 

Figure~\ref{fig:figure_3} shows a transition from regular to chaotic when the mass ratio changes from $\kappa=1$ to larger values. The degeneracies in the energy spectrum for $\kappa=1$ and $1/\kappa=0$ lead to well spaced bins in the figure. This behavior is however rather unique, and it is immediately broken for other mass ratios. For example, already for $\kappa=1.2$ the levels start to repel each other, and the distribution $P(s)$ can be approximated by the Wigner distribution~\cite{Stockmann1999}
\begin{equation}
P_{GOE}(s)=\frac{\pi s}{2}e^{-\frac{\pi s^2}{4}}.
\label{eq:p_goe}
\end{equation}
Note that it is important to use only one value of $p$ for this conclusion. Levels that correspond to different values of $p$ do not repel each other, and the Wigner distribution cannot be realized~\cite{Kaufman1999}.

\begin{figure}
\includegraphics[scale=0.55]{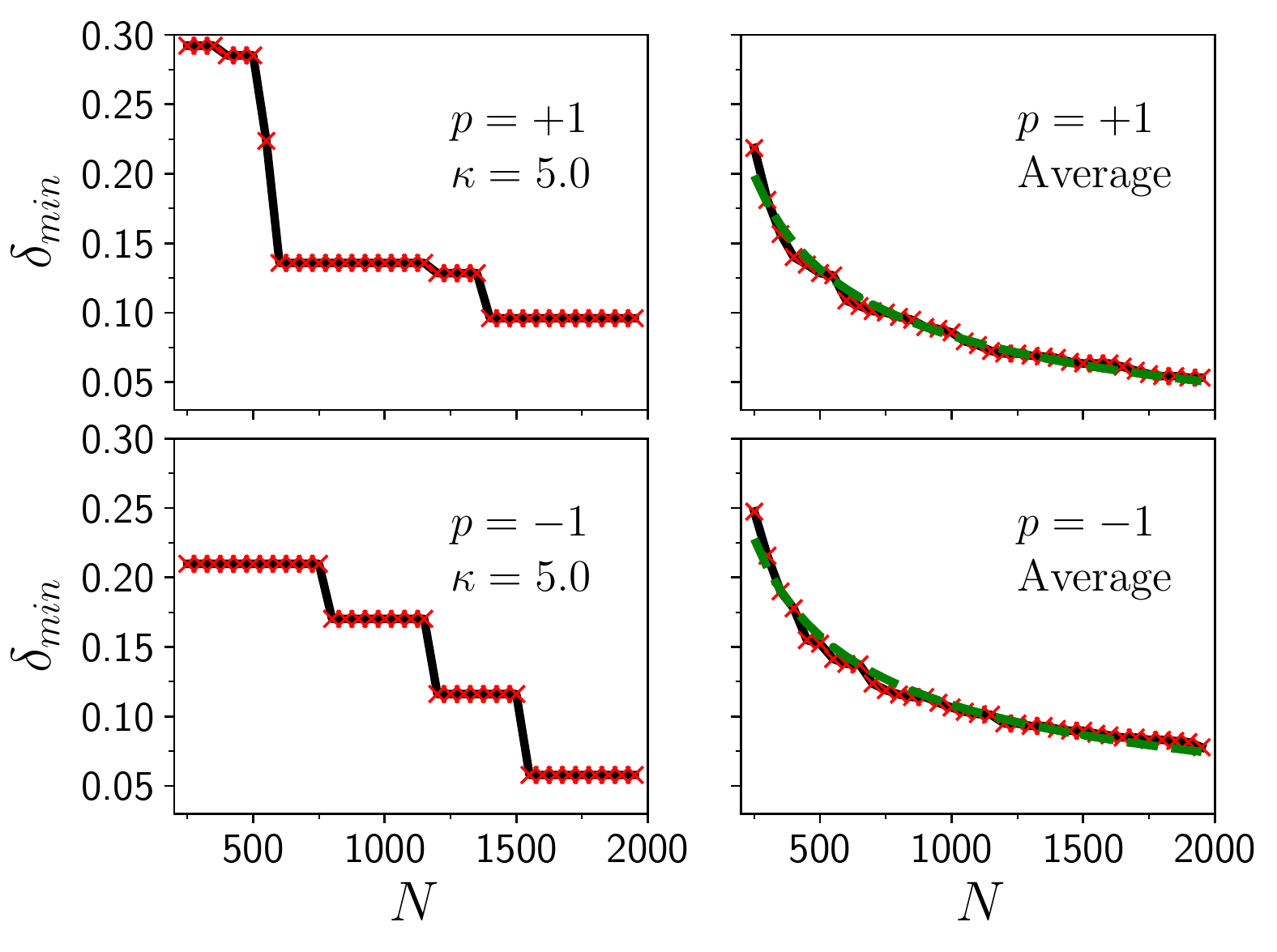}
\caption{The minimal distance between energy levels as a function of the number of considered  levels. The data points are given as (red) crosses. The (black) solid curve is added to guide the eye. The left panels show $\delta_{min}$ for $\kappa=5$ and $p=\pm 1$. The right panels display $\delta_{min}$ averaged over different mass ratios $\kappa$ (see the text for details). The (green) dashed curves in the right panels show the best-fit to the asymptotic $1/\sqrt{N}$ behavior, which is expected for random matrices.}
\label{fig:figure_4}
\end{figure}

It is impossible to analyze every value of $\kappa$. However, we can also say something on average about our system. 
To that end, we calculate the dependence of the minimal distance between levels as a function of the number of considered  levels, $\delta_{min}(N)=\mathrm{inf}\{E_n-E_{n-1}\}_{n<N}$, here $E_n$ is the energy of the $n$th state with a given value of $p$. For a random matrix, $\delta_{min}$ is expected to scale as $1/\sqrt{N}$~\cite{Benarous2013,Blomer2017}.
To the best of our knowledge, this result is not strictly proven, but at the intuitive level it can be understood from Eq.~(\ref{eq:p_goe}). The probability that the distance between two energy levels is smaller than $\delta_{min}$ is given by $\int_0^{\delta_{min}} P_{GOE}\mathrm{d}s=1-e^{-\pi \delta_{min}^2/4}$. In the limit $\delta_{min}\to 0$, this expression can be approximated by $\pi \delta_{min}^2/4$. If we consider $N$ lowest states, then the probability that all nearest neighbors are separated by $\delta>\delta_{min}$ is given by 
$(1-\pi \delta_{min}^2/4)^N$. To keep this probability independent of $N$, the parameter $\delta_{min}$ must be proportional to $1/\sqrt{N}$.  

We show $\delta_{min}$ for our system for $\kappa=5$ in Fig.~\ref{fig:figure_4} (left panels). We see that for a given value of $\kappa$ it is impossible to verify $1/\sqrt{N}$ scaling, at least for the considered amount of eigenstates. However, the randomness present in a mass-imbalanced system can be recovered. To show this, we average $\delta_{min}$ over different masses, i.e., $\delta^{average}_{min}=\frac{1}{\cal{M}}\sum \delta_{min}(\kappa_i)$, where $\cal{M}$ determines how many values of $\kappa$ appear in the sum. To produce Fig.~\ref{fig:figure_4} (right panels), we sum over the following values of the mass ratios: $\kappa=1.1,1.2,...,5$.   The parameter $\delta^{average}_{min}$ has approximately $1/\sqrt{N}$ behavior at large values of $N$, which confirms our expectation that systems with $1/\kappa\in (0,1)$ are not integrable.

\subsection{Wave function}

\begin{figure}[t]
\vspace*{0.2cm}
\includegraphics[scale=0.5]{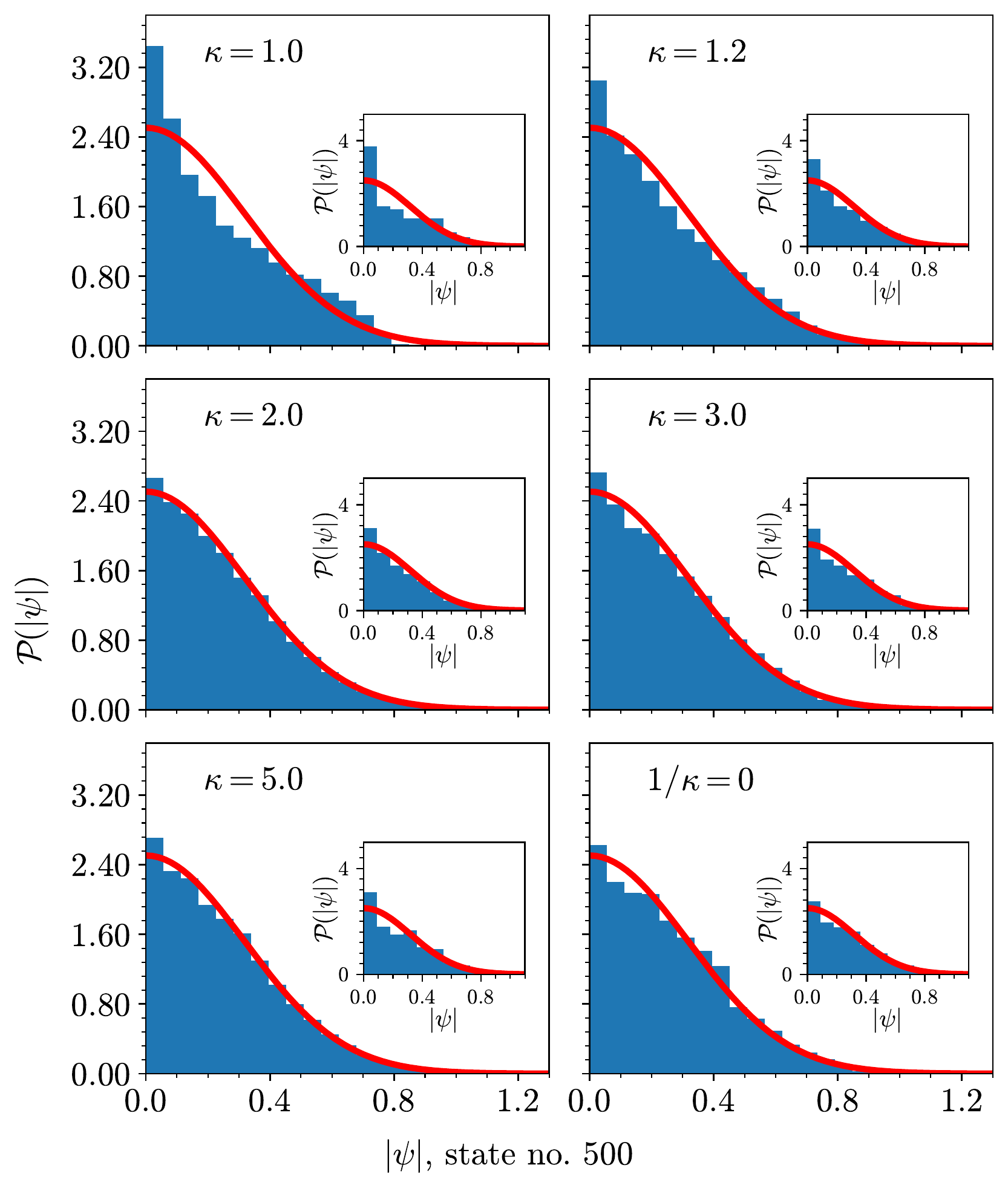}
\caption{The distributions of values of $|\psi|$ for different mass ratios, $\kappa=m_I/m$. The histogram describes the probability that a pixel has a value of $|\psi|$ in a specific interval. For the main plots, we use a $315\times 315$ pixel raster image of $|\psi|$. 
 The insets are showing $\mathcal{P}(\psi)$ for the state represented by $33\times 33$ pixel image. The (red) solid curve shows the prediction of Eq.~(\ref{eq:p_psi}) (no fit parameters).}
\label{fig:figure_5}
\end{figure}

The analysis above shows a drastic change of properties of the system when moving from integrable to non-integrable regimes. Information about this transition is extracted by analyzing the energy levels as in Fig.~\ref{fig:figure_3}, although, the correspondence principle conjectures properties at the level of individual wave functions. The wave function of a highly excited state contains too much information for the human eye, and one has to rely on a few existing conjectures that allow one to connect classical chaos to quantum states.
For example, the chaotic states are expected to be similar to a random superposition of plane waves~\cite{Berry1977b}, since the underlying classical phase space has no structure, i.e., the classical motion is not associated with motion on a two-dimensional torus. This expectation applies to a typical random state (not to atypical, e.g., scared states~\cite{Kaplan1998}). 
In contrast, the wave functions of integrable states are expected to have some non-trivial morphology, since classical phase space of integrable systems has some structure. Below, we illustrate these ideas for our problem. We focus on a distribution of wave-function amplitudes, although, 
other signatures of `chaos' in eigenstates connected to local currents and nodal lines\footnote{A nodal line is a set of points $\{X,Y\}$ that satisfy $\psi(X,Y)=0$.}~\cite{Evans1998,Berggren1999, Jain2017} will also be important when we analyze our neural network.

A celebrated result of the random-wave conjecture is a Gaussian distribution of wave-function amplitudes, see examples in Refs.~\cite{Shapiro1984, McDonald1988,Samajdar2018}:
\begin{equation}
\mathcal{P}(\psi)=\frac{1}{\sqrt{2\pi}v}e^{-\frac{|\psi|^2}{2v^2}},
\label{eq:p_psi}
\end{equation} 
where the variance $v=1/L$ fixes the normalization of the wave function\footnote{The value of $v$ is calculated by using the average value of $\psi^2$, i.e., $\overline\psi^2=\int x^2P(x)\mathrm{d}x$, in the normalization condition, i.e.,$\int \psi^2\mathrm{d}z_1\mathrm{d}z_2=\overline\psi^2\int \mathrm{d}z_1\mathrm{d}z_2=1$, which leads to the condition on $v$ as $\int x^2P(x)\mathrm{d}x=1/\int\mathrm{d}z_1\mathrm{d}z_2$.}.
We present our numerical calculations of $\mathcal{P}(\psi)$ in Fig.~\ref{fig:figure_5}. For this figure, we discretize the wave function for the 500th state using either a $315\times 315$ pixel grid or a $33\times 33$ pixel grid, and assign to each unit of the grid a value that corresponds to $\psi$ in the center of the unit. The distribution of these central values for a given value of $\kappa$ is presented as a histogram in Fig.~\ref{fig:figure_5}.
For $\kappa=1$ the $\mathcal{P}(\psi)$ resembles an exponential function (cf.~Ref.~\cite{Samajdar2018}). For larger values of $\kappa$, a Gaussian profile is developed.
The distinction between the histogram and Eq.~(\ref{eq:p_psi}) is clear for $\kappa=1$.  For $1/\kappa=0$ the difference is less evident. Note that the peak at $\psi=0$ is enhanced in comparison to the prediction of Eq.~(\ref{eq:p_psi}) for all values of $\kappa$. This is due to the evanescence of the wave function at the boundaries, which is a finite-size effect beyond Eq.~(\ref{eq:p_psi}). Finally, the characteristics of the states are also visible in a low-resolution images, see the insets of Fig.~\ref{fig:figure_5}. This feature will be used in the design of our neural network discussed below.

\section{Neural Network}
\label{sec:network}

To construct a neural network that can distinguish integrable states from non-integrable ones,
we need to

\begin{itemize}
\item[A.] prepare a data set for training the network
\item[B.] choose a suitable architecture and a training algorithm
\end{itemize}
In this section, we discuss these two items in detail.

\subsection{A data set}
As data set we use the set $\mathcal{A}$ made of two-dimensional images that represent highly excited states. We can use images of (real) wave functions, $\psi(z_1,z_2)$ or probability densities, $|\psi(z_1,z_2)|^2$. We have checked that these two representations lead to similar results. In the paper, we present only our findings for $|\psi(z_1,z_2)|^2$.    
To produce $\mathcal{A}$, we diagonalize the Hamiltonian $H_{P=0}$ of Eq.~(\ref{eq:h_p}) for $\kappa=1,2,5$ and $1/\kappa=0$. 
Each image has a label -- integrable (for $\kappa=1$ and $1/\kappa=0$) or non-integrable (for $\kappa=2$ and $5$).\footnote{To avoid any bias towards non-integrable states, we use non-integrable states for only two values of $\kappa$. However, we have checked that our conclusions hold also true if we include other values of $\kappa$ into the data set, in particular, if we add 1000 states to $\mathcal{A}$ from a system with $\kappa=15$.} We do not include information about the mirror symmetry, i.e., states with different values of $p$ are treated on the same footing, since we do not expect that this information is relevant for a coarse-grained (see below) image of $|\psi(z_1,z_2)|^2$. This allows us to work with twice as large datasets compared to Fig.~(\ref{fig:figure_3}).   Each mass ratio contributes 1000 states to $\mathcal{A}$, which therefore contains 4000 images in total. It is reasonable to not use data sets that contain states with very different energies: Very different energies lead to very different length scales, and hence different information content that should be learned. We choose to include all states from the 50th to 1050th excited states. Not much should be deducible about the low-lying states (with $N\sim 10$) from the correspondence principle, therefore, we do not use them in our study.

\begin{figure}[t]
\includegraphics[scale=0.8]{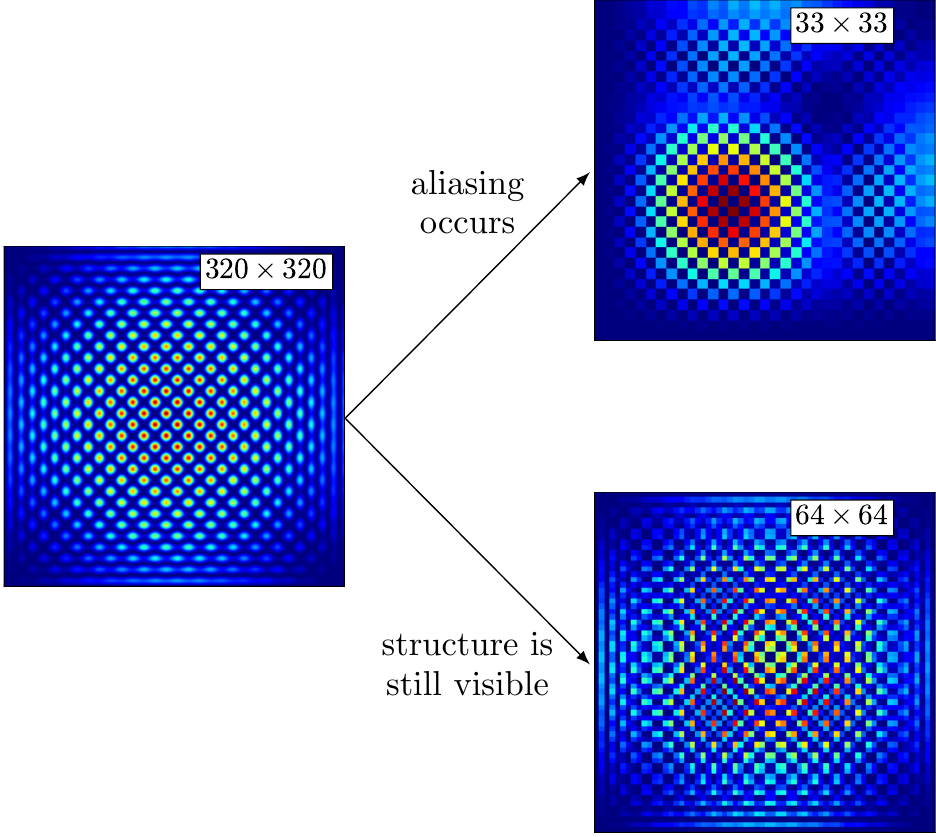}
\caption{Schematic representations of a probability distribution of a state for $1/\kappa=0$ susceptible to aliasing. The left image shows a high-resolution representation ($320\times 320$ pixel image). This representation contains too many pixels for our purposes, and can be optimized. The right images present two low-resolution representations, which we use to train the network. The representation with $33\times 33$ pixels does not contain enough information and spatial aliasing occurs. The representation with $64\times 64$ pixels contains all relevant information, and is used for the analysis in Sec.~\ref{sec:results}.}
\label{fig:figure_6}
\end{figure}

A wave function $\psi(z_1,z_2)$ is a continuous function of the variables $z_1$ and $z_2$, see Fig.~\ref{fig:figure_1} (b). To use it as an input for a network, we need to  discretize  and coarse-grain it. To this end, we represent $\psi$ as a $64 \times 64$ pixel image, and as the value of the 
pixel we use the value of the wave function at the center of the pixel.\footnote{The color depth (i.e., how many colors are available) of a pixel is effectively given by the numerical precision used to produce the input data. If experimental data is used as an input, then their accuracy will determine the color depth of a pixel.} The resolution is important for this discretization. Low resolution might not be able to capture oscillations present in highly excited states, leading to a loss of important physical information. For example, the approximately $N$th state in the spectrum for $1/\kappa=0$ will have about $\sqrt{N}$ oscillations in each direction, and it is important therefore to use a $2 \sqrt{N}\times 2\sqrt{N}$ representation of the wave function (similar to the Nyquist–Shannon sampling theorem)\footnote{It is worth noting that a general (in particular, non-integrable) $N$th state might have fewer nodal domains [a nodal domain is a domain bounded by a nodal line]~\cite{Jain2017}, and allow for a representation with a smaller number of pixels. However, a general data set may have states with as many as $N$ nodal domains (cf. Courant's theorem), which requires a $2 \sqrt{N}\times 2\sqrt{N}$ representation of the wave function.}. For a lower resolution,  the  oscillations are not faithfully reproduced in the low resolution image and spatial aliasing occurs.
We illustrate this using  the $33\times 33$ resolution in Fig.~\ref{fig:figure_6} for an integrable state that is susceptible to spatial aliasing.

 Note that out of curiosity, we have also used images with $33\times 33$ pixel resolution to train our network. The network could reach relatively high accuracy (higher than $90\%$). However, not all integrable states were detected properly. For example, the one in Fig.~\ref{fig:figure_6} was classified as non-integrable by the network. In general, spatial aliasing is more damaging for integrable states, which have symmetries that should be respected; non-integrable states are more random, and some noise does not change the classification of the network. Everywhere below we use the $64\times 64$ pixel representation, which gives a sufficiently accurate representation of the state, so that we do not need to worry that the network learns unphysical properties. Note that certain local features (e.g., $\psi(z_1=z_2)=0$) of the wave function may disappear at this resolution. The overall high accuracy of our network suggests that such features are not important for our analysis.

The set $\mathcal{A}$ seems somewhat small. For example, the well-known
Asirra dataset~\cite{Asirra}
for the cat-dog classification contains 25000 images that are commonly used to test and compare different image recognition techniques. 
However, we will see that $\mathcal{A}$ is large enough to train a network that can accurately classify integrable and non-integrable states. The dataset $\mathcal{A}$ is further divided into two parts. We draw randomly 85~\% of all states and use them as a training set. The remaining 15~\% is used for testing. We fix the random seed used for drawing to avoid discrepancies between different realizations of the network.  It is worthwhile noting that in image-recognition applications, the dataset $\mathcal{A}$ may be divided into more than two parts. For example, in addition to the training set and testing set, one can introduce a validation (or development) set~\cite{James2013}, which is used to fine-tune parameters of the model. We do not use this additional set here. The focus of this work is on understanding features of our general image classificator, and not on improving its accuracy.

\subsection{Architecture}
The neural network in our problem is a map that acts in the space $\mathcal{X}$ made of all $64\times 64$ pixel representations of $|\psi|^2$. By analogy to the standard dog-vs-cat classifier, the output of the network is a vector with two elements $\mathbf{b}$. Note that $n$ output neurons are usual for classifying $n$ classes. However, it is possible to use a single output neuron for a binary classification, since we know that $b_1+b_2$ must be equal to one. We use two neurons, since such an architecture can be straightforwardly extended to more output neurons, which may be useful for the future studies, as we discuss in Sec.~\ref{sec:concl}.

The first element of the output layer,  $0\leq b_1 \leq 1$, determines the probability that the input state is integrable, whereas the second element $b_2=1-b_1$ is the probability that the input state is non-integrable. An input state is classified as integrable (non-integrable) if $b_1>b_2$ ($b_1<b_2$). 

Mathematically, the network is a map $\mathbf{f}$, which acts on the element $a$ of $\mathcal{X}$ as
\begin{equation}
\mathbf{f}(a;\theta, \theta_{hyp})=\mathbf{b}.
\label{eq:mapping_network}
\end{equation}
The $\mathbf{f}$ is determined by the set of parameters $\theta$, which are optimized by training. Since our problem is similar to image recognition (in particular dog-vs-cat classification)~\cite{LeCun2015}, which is one of the standard applications in machine learning, we can use the already known training routines (SGD, ADAM, Adadelta,...) for optimizing $\theta$. The outcome for the parameters $\theta$ may vary between different trainings, and we use this variability to check the universality of our results. Specifications of $\mathbf{f}$ that are not trained but specified by the user are called hyperparameters ($\theta_{hyp}$). Examples of them include the loss function, optimization algorithm, learning rate, network architecture, size of batches that the data is split into for training (batch size) and the length of training (epochs). We find hyperparameters by trial-and-error. 

\begin{widetext}

\begin{figure}
\includegraphics[scale=0.9]{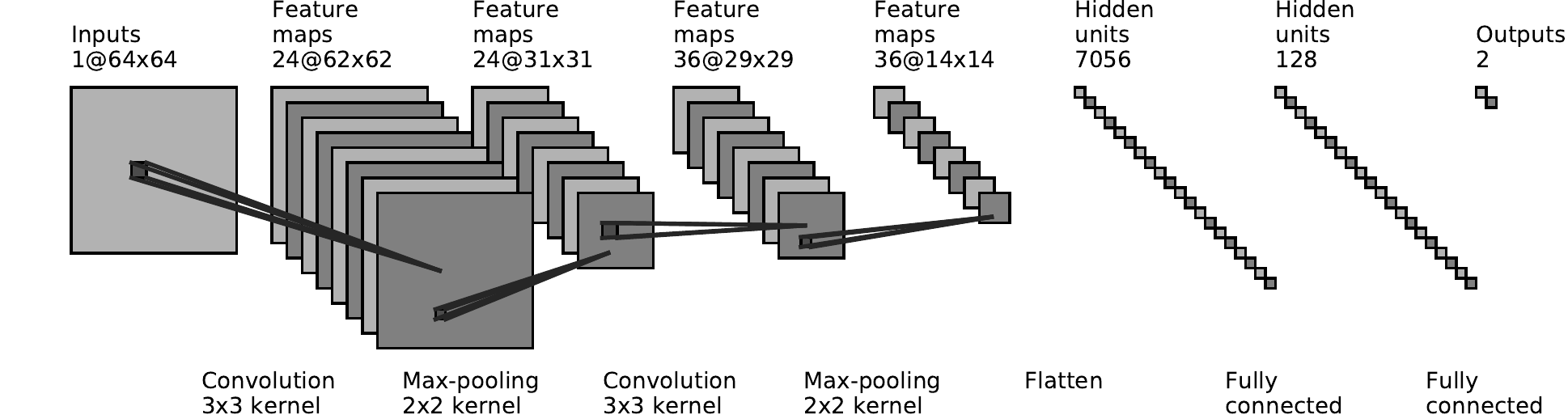}
\caption{An illustration of the ConvNet used in our analysis. An input layer, $64\times 64$ image, is followed by a sequence of layers: a convolutional layer, a pooling layer, a convolutional layer, a pooling layer, and two fully connected layers. The last layer is used to produce an output layer, which is made out of two neurons. In this neural network, we use the rectified linear activation function~(ReLU).}
\label{fig:figure_7}
\end{figure}

\end{widetext}

The simplest form of a network is called a dense network in which all input neurons are connected to all output neurons. However in most cases of image detection, this architecture does not lead to accurate results. This also happens in our case. Instead, we resort to a standard architecture based on ConvNets for image recognition, see Fig.~\ref{fig:figure_7}.\footnote{This figure is generated by adapting the code from \url{https://github.com/gwding/draw_convnet}~.} 
Our network consists of two convolutional layers and two max-pooling layers. 
The former use a set of filters and apply them in parts to the image to produce
a new smaller image. This is somewhat analogous to a renormalization group transformation~\cite{Wilson1983}. A set of images that are produced by a convolutional layer is called feature map.  Each convolutional layer is followed by a max-pooling layer which reduces the size of an image. The size of max-pooling layers is a hyperparameter. In our implementation,
max-pooling layers take the largest pixel out of groups of 2 by 2. 

One could use architectures different from the one presented in Fig.~\ref{fig:figure_7}. However, we checked that they do not lead to noticeably different results. Therefore, we do not investigate this possibility further.

\section{Numerical Experiments}
\label{sec:results}

\begin{figure}[t]
\includegraphics[scale=0.35]{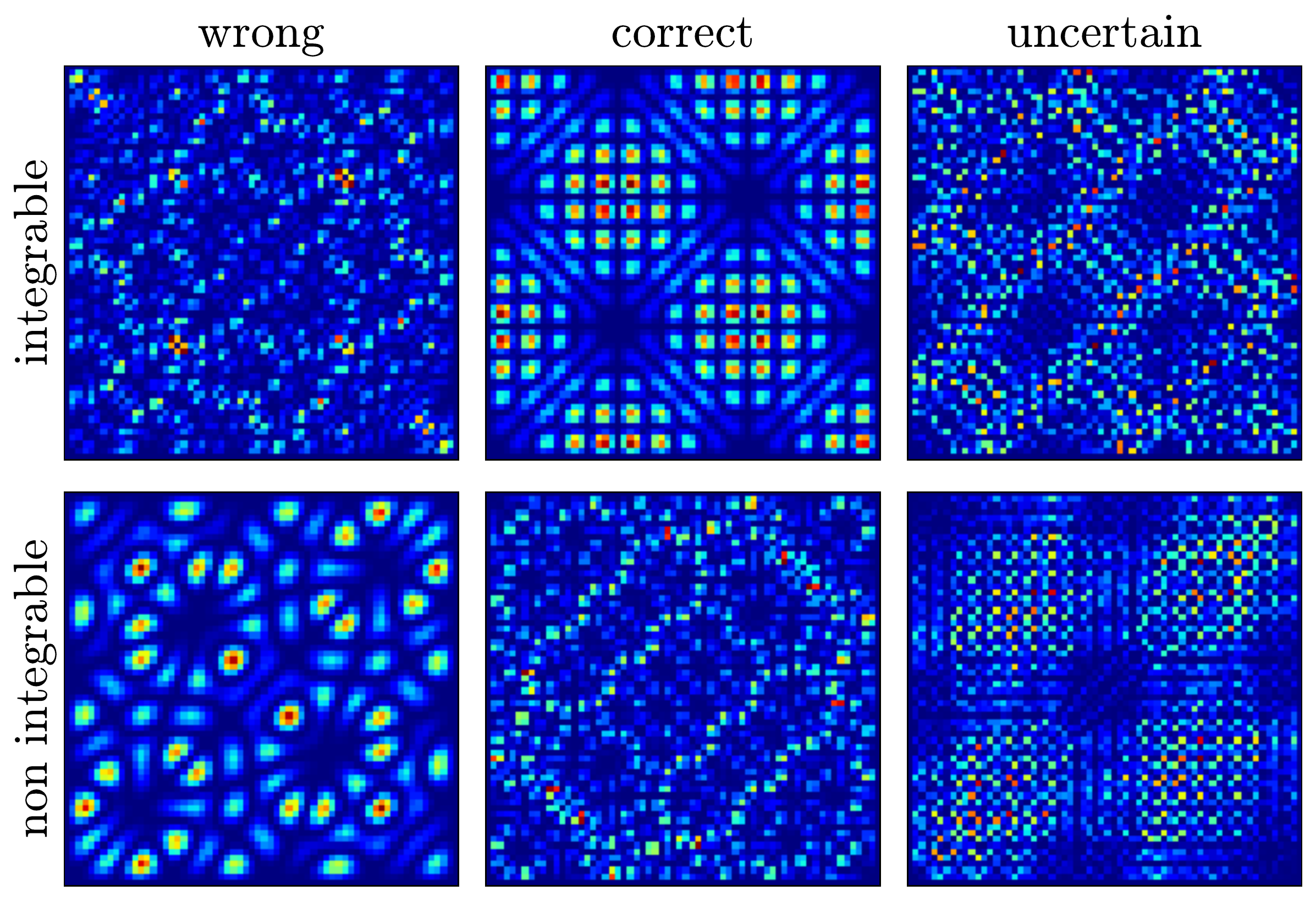}
\caption{The figure shows $|\psi|^2$ of exemplary integrable (upper row) and non-integrable states (lower row) together with the corresponding prediction of the network. The network assigns a wrong (correct) label for the states in the first (second) column. The third column shows states, which are identified correctly by the network, but with a low confidence level (with about $60\%$). In other words, the states in the third column confuse the network and lead to $b_1\simeq b_2$ in Eq.~(\ref{eq:mapping_network}).}
\label{fig:figure_8}
\end{figure}

Following the discussion in Sec.~\ref{sec:network}, we train and test the neural network. We observe that a typical accuracy of the trained network (which we refer later to as $\mathcal{N}$) is $\sim 97\%$.\footnote{We use the word `typical' to emphasize that a trained network depends on hyperparameters and random seeds. Even for a given set of hyperparameters, each set of random parameters leads to a slightly different network $\mathcal{N}$. We can tune hyperparameters to reach higher accuracies. We do not discuss this possibility here, since high accuracy is not the main purpose of our study.}
This means that about 18 states out of 600 used for testing are given the wrong label. Out of these 18 states, roughly one half is integrable. We illustrate typical wave functions that are classified correctly and wrongly in Fig.~\ref{fig:figure_8}. 
It does not mean that these states are in anyway special -- another implementation (e.g., another random seed for weights) will lead to other states that are given the wrong label. Non-integrable states with some structure (e.g., states with scars) in general confuse the network and might be classified as integrable.

\begin{figure}[t]
\includegraphics[scale=0.45]{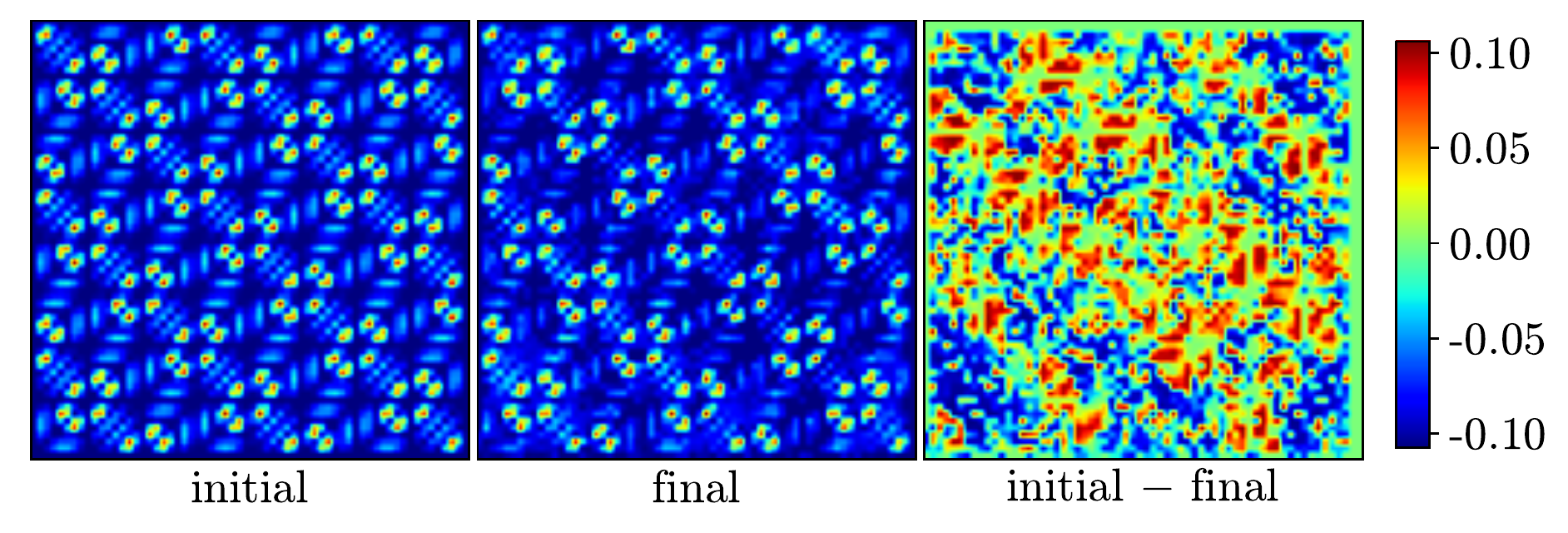}
\includegraphics[scale=0.45]{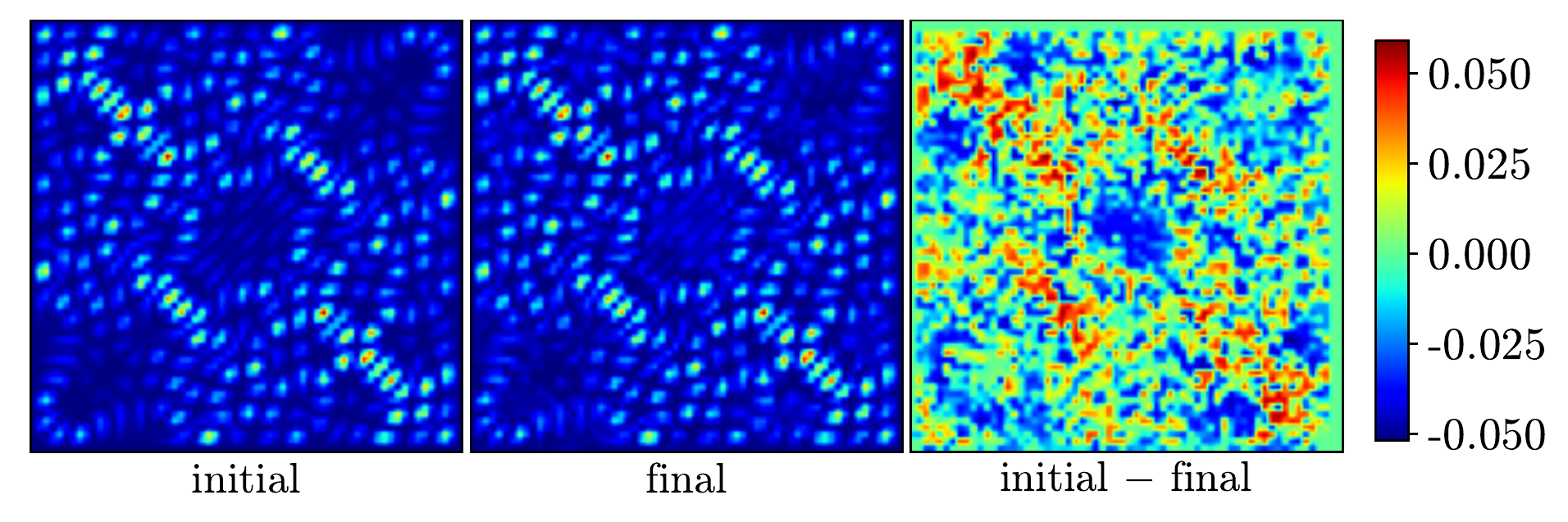}
\caption{Fooling the network. The first column shows $|\psi|^2$ of the initial state, which is correctly identified by the network as integrable (non-integrable) in the upper (lower) panel. The second column shows a slightly modified image of $|\psi|^2$, which is wrongly identified by the network. 
The third column shows the difference between the first and second columns divided by the maximum value of the initial state. The color chart corresponds to the images in the third graph only. }
\label{fig:figure_12}
\end{figure}

In general, it is hard to interpret predictions of the neural network. 
This becomes clear after noticing that some images can be changed so that a human eye can hardly detect any variation. At the same time, this change completely modifies the prediction of the network. Such a change can be accomplished especially easily for integrable states\footnote{Tools of adversarial machine learning can make a neural network classify an arbitrary integrable state as `non-integrable' using a small number ($\sim 100$) of iterations.  This is not possible for an arbitrary non-integrable state.}, thus, deep learning (DL)  confirms our intuition that integrable states are a small subset in the space of all possible states. However, such a situation can also occur for non-integrable (in particular scared) states. We illustrate this in Fig.~\ref{fig:figure_12}, which is obtained by slightly modifying states from $\mathcal{A}$ using tools of adversarial machine learning, see Ref.~\cite{szegedy2014intriguing}.

One simple way to extract features of the network is to look at feature maps, which should contain information about what features are important. For example, the first layer might represent edges at particular parts of the image, the second might detect specific arrangements of edges, etc. 
However, we could not extract any meaningful information from this analysis. This is expected: the features of integrable and non-integrable states are more abstract and not as intuitive as the features of cats and dogs or images of other objects we encounter in everyday life.

\begin{figure}[t]
\includegraphics[scale=0.55]{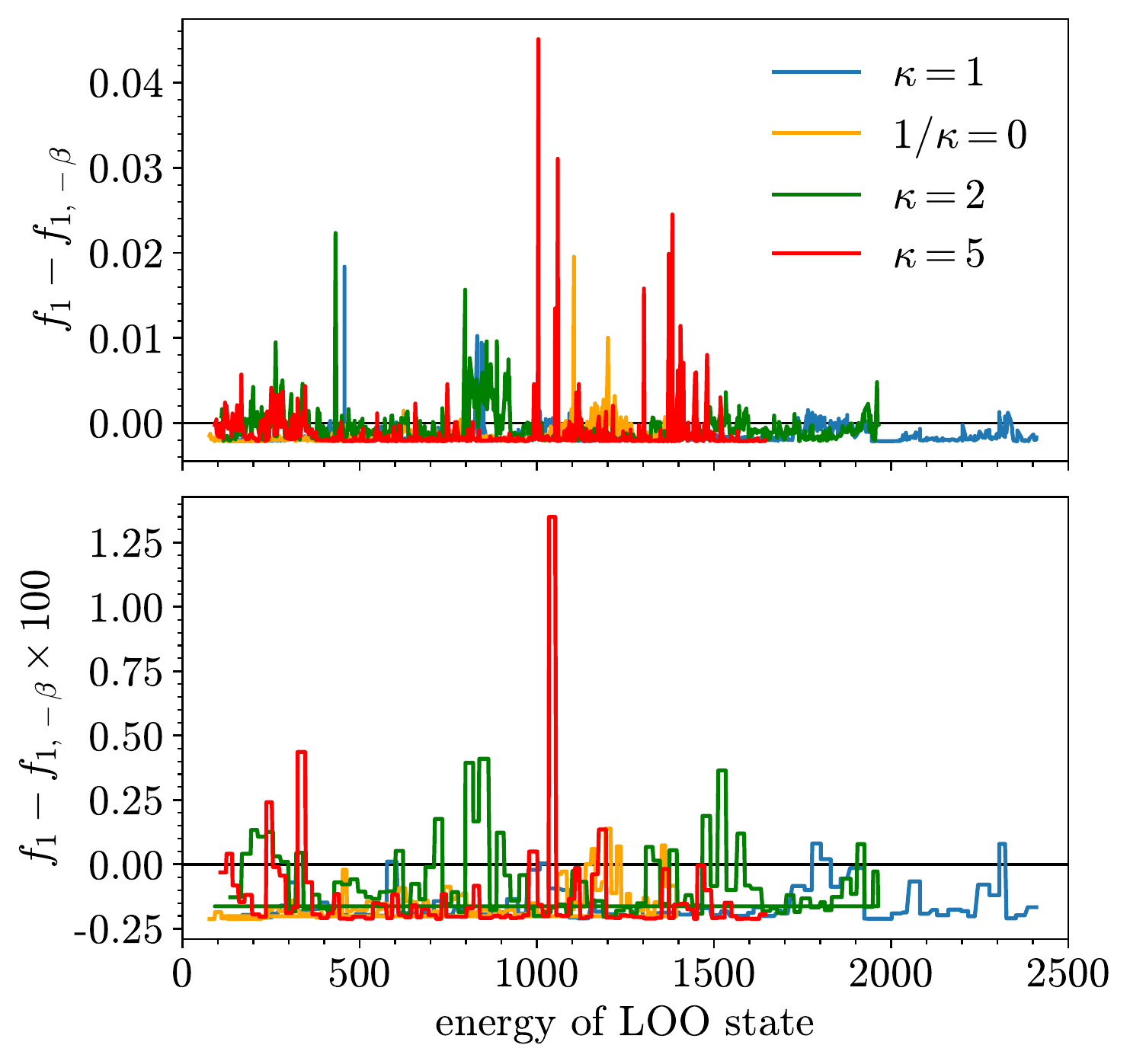}
\caption{A typical outcome of a leave-one-out (LOO) algorithm. (Top)  The panel shows $f_1-f_{1,-\beta}$ for $\beta$ that consists of a single state as a function of the energy of the state that was left out. (Bottom) The panel shows $f_1-f_{1,-\beta}$ for $\beta$ that consists of ten consecutive states as a function of the energy of the first state in $\beta$. The test state, $\psi$, here is for $\kappa=5$, its energy is $534.86$ (in our units).  }
\label{fig:figure_LOO}
\end{figure}

Other approaches to analyze a network rely on estimating the effect of removing a single (or a group) of elements on a model. For large data sets, this can be done by introducing influence functions~\cite{Hampel1974, Koh2017}. Here, we work with a small data set, and, therefore, we can calculate directly the actual effect of leaving states out of training on a given prediction. Our goal is to understand correlations between states of different energies. 
In our implementation we compare the prediction of $\mathbf{f}$
from Eq.~(\ref{eq:mapping_network})
for $|\psi|^2$ to a prediction of $\mathbf{f}_{-\beta}$ for the same state. Here $\mathbf{f}_{-\beta}$ is obtained by training a neural network after leaving out the set $\beta$ from $\mathcal{A}$. The comparison of the two predictions ($\mathbf{f}-\mathbf{f}_{-\beta}$) allows us to estimate the importance of the set $\beta$ for the classification of a test state $\psi$.\footnote{Note that it is important to choose a test state $\psi$ for which the network gives an accurate prediction with high confidence level, i.e., $b_i\to 1$. For other states, an intrinsic randomness of ConvNets can lead to a drastic change in the classification of the network.}    We present a typical example in Fig.~\ref{fig:figure_LOO}, where one observes no clear energy correlations, which suggests that the network learns different energies simultaneously, at least in the energy interval we choose to work with. This observation is consistent with our discussion below that the network does not learn specific features of non-integrable states, and only overall randomness, which does not depend on a specific energy window. 
Finally, we note that our result in Fig.~\ref{fig:figure_LOO} is an example of cross-validation. It suggests weak dependence of the output of the network on a particular state, which is a necessary condition for a good performance of our neural network.

Below, we explore the network $\mathcal{N}$ further. To this end, we resort to numerical experiments. We employ $\mathcal{N}$ to analyze states outside of the set $\mathcal{A}$. First, we study physical states, and then non-physical ones.

\subsection{Classification of physical states outside of $\mathcal{A}$.}
\begin{figure}[t]
\includegraphics[scale=0.6]{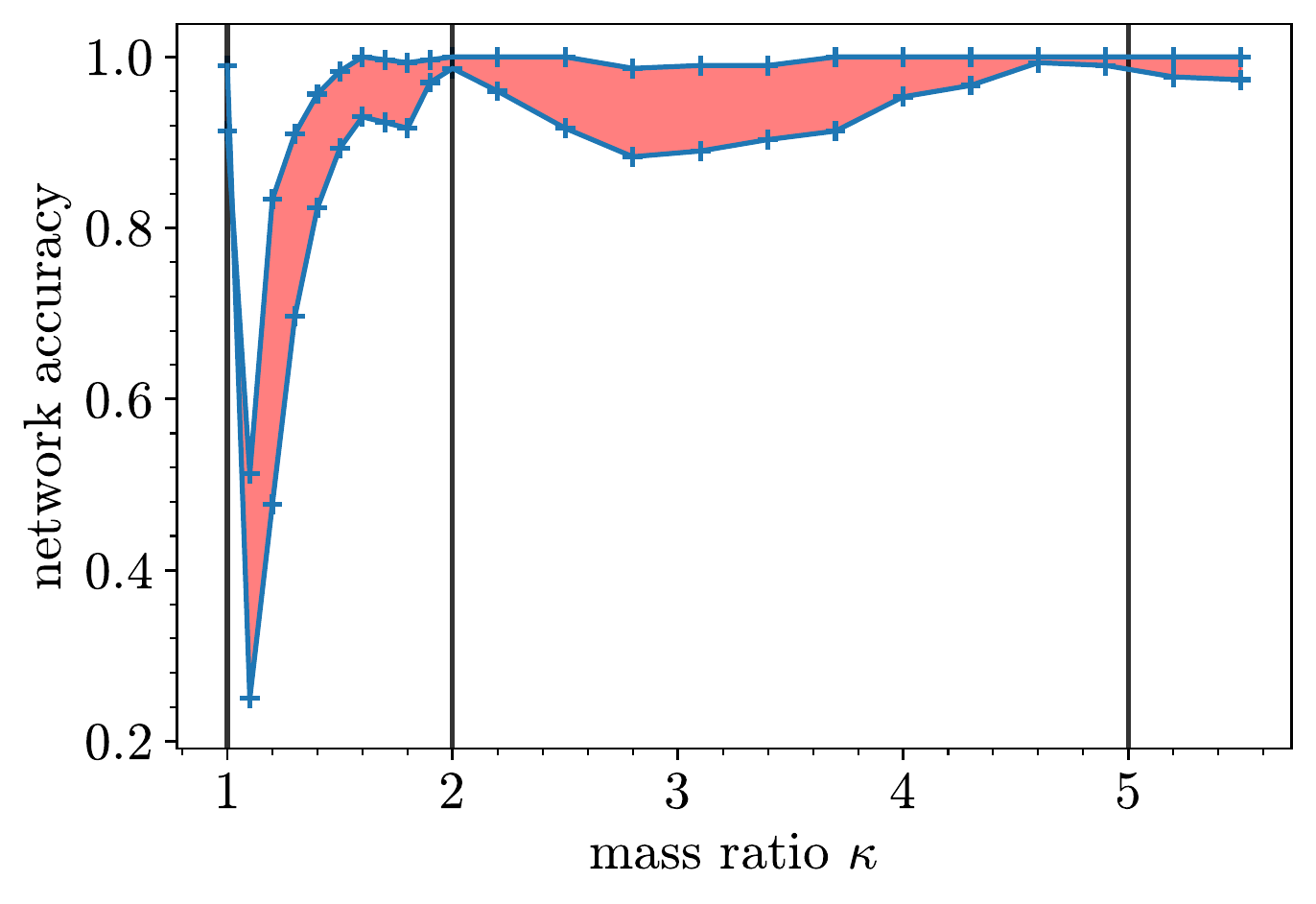}
\caption{Accuracy of the network $\mathcal{N}$ as a function of the mass imbalance $\kappa$. The shaded area shows the uncertainty area. To obtain it, we train a set of neural networks using different random seeds. The highest (lowest) accuracy from a set produces an upper (lower) bound of the uncertainty area. The dots are calculated, and the curves are added to guide the eye.  The points $\kappa=1,2,5$, which are used for training, are shown with vertical lines. The point $\kappa=1$ corresponds to an integrable system. All other points of the $x$-axis are expected to be non-integrable (cf.~Fig.~\ref{fig:figure_3}).}
\label{fig:figure_9}
\end{figure}

As a first application of $\mathcal{N}$, we use it to classify eigen-states of $H_{P=0}$ not used in the training, i.e., for $\kappa\neq 1,2,5$ and $1/\kappa\neq 0$. These states are non-integrable (cf.~Fig.~\ref{fig:figure_4}), and we observe that $\mathcal{N}$ accurately classifies them as such as long as $\kappa$ is far enough from $\kappa=1$, see Fig.~\ref{fig:figure_9}. 
The figure shows that predictions of $\mathcal{N}$ are not accurate only for systems with $\kappa=1+\epsilon$, where $\epsilon$ is a small parameter. These systems are non-integrable, however, the morphology of their eigenstates is very similar to the integrable ones at $\kappa=1$. The network classifies them wrongly because of this. Already for $\kappa\simeq 1.5$, the accuracy of the network is close to one, and it stays high for larger values of $\kappa$. The region between $1$ and $1.5$ can be interpreted as a transition of the network classification from integrable to chaotic~\cite{Kharkov2020}. We do not expect this region to be universal -- it should depend on hyperparameters, and the states used for training of $\mathcal{N}$. Therefore, we do not investigate it further.

To test the network on integrable states, we use wave functions of two non-interacting bosons in a box potential of size $L$:
\begin{equation}
\Psi_{\mathrm{B}}=\frac{N_{k_1,k_2}}{L}\left[\sin(k_1z_1)\sin(k_2z_2)+\sin(k_2z_1)\sin(k_1z_2)\right],
\end{equation}
where $k_1\leq k_2$, and $N_{k_1,k_2}$ is a normalization constant, $N_{k_1=k_2}=1$ and $N_{k_1<k_2}=\sqrt{2}$.  The order of the states is determined by the energy $(k_1^2+k_2^2)/2$. The set of functions $\{\Psi_{\mathrm{B}}\}$ is complementary to the $1/\kappa=0$ case studied above for fermions. The bosonic symmetry yields states that are orthogonal to the training set, and therefore, the training routine can have no microscopic information about the wave function, $\Psi_{\mathrm{B}}$. We use 1000 states of the bosonic type (from the 50th to 1050th) as an input for $\mathcal{N}$.  We observe that $\mathcal{N}$ accurately (accuracy is $\simeq 96\%$) classifies states $\Psi_{\mathrm{B}}$ as integrable. 

To connect the analysis of $\Psi_{\mathrm{B}}$ to studies of quantum billiards, we note that two bosonic impurities in an infinite square well can be mapped on a right triangle with two impenetrable boundaries. At the third boundary a zero Neumann boundary condition should be satisfied -- $\Psi_B'|_{z_1=z_2}=0$. The mapping follows from the mapping discussed for fermions~(see Fig.~\ref{fig:figure_1}) assuming that the impurity is infinitely heavy. In particular, Fig.~\ref{fig:figure_1}~b) shows the geometry of the problem in this case. Note that the bosonic symmetry requires that the derivative of the wave function vanishes on the diagonal of the square in Fig.~\ref{fig:figure_1}~b). The high accuracy of the classification of the bosonic states suggests that a network trained using the Dirichlet boundary condition can also be used to classify states with the Neumann boundary condition. In other words, the network is mainly concerned with the `bulk' properties of the wave function, the boundary is not important.

\subsection{Classification of non-physical states}
\label{sec:unphys}

\begin{figure}[t]
\includegraphics[scale=0.6]{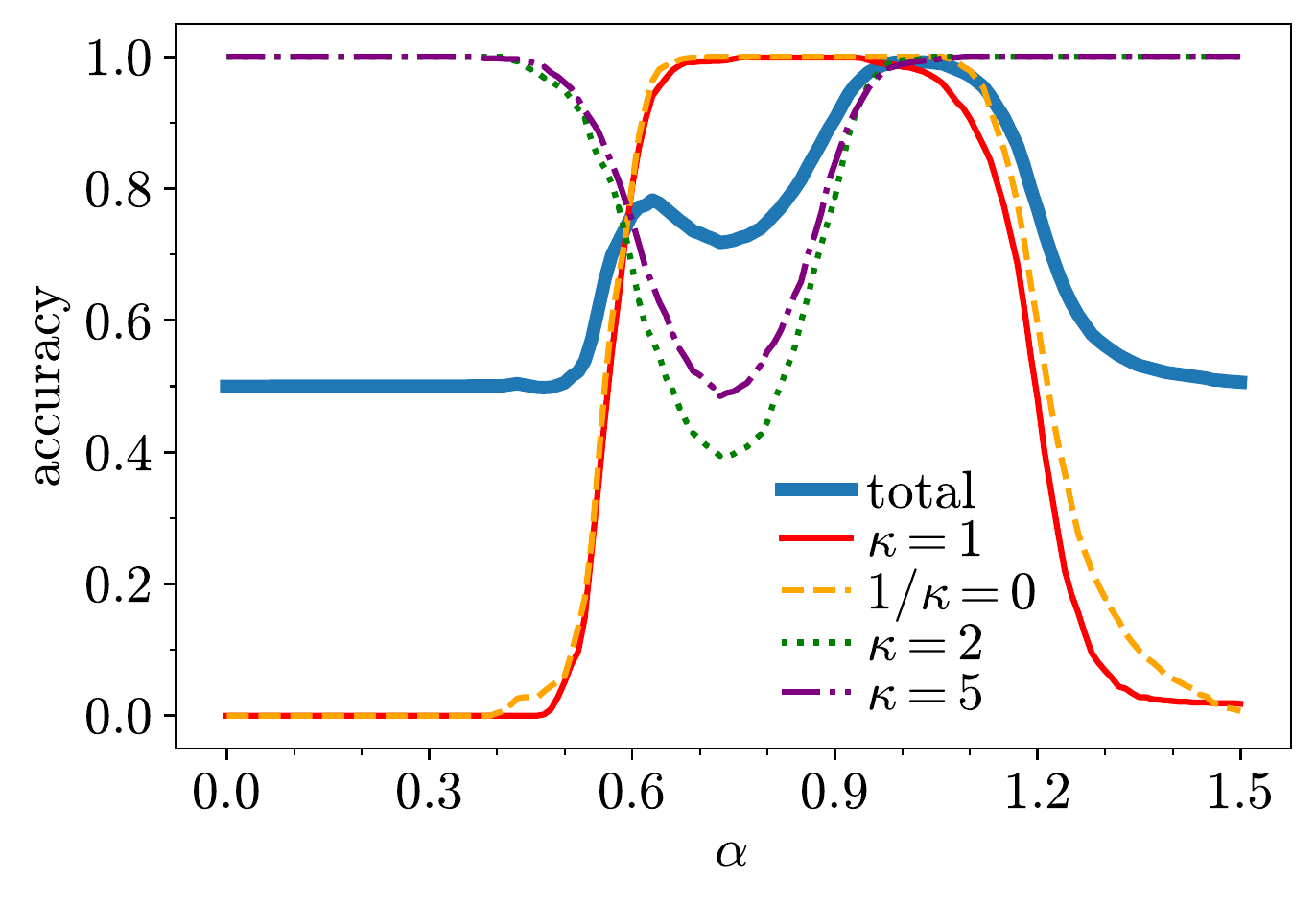}
\caption{Prediction of the network for the states from $\mathcal{A}$ multiplied by a factor $\alpha$.
The curves show the accuracies for four values of the mass ratio~$\kappa$. The average of these four curves is shown as a thick solid curve.}
\label{fig:figure_10}
\end{figure}

The network $\mathcal{N}$ can classify any $64\times 64$ pixel input image, and it is interesting to explore the outcome of the network for images that have no direct physical meaning. We start by considering non-normalized eigenstates of $H_{P=0}$. The normalization coefficient does not change the physics behind the states. However, since the function $\mathbf{f}$ is non-linear, i.e., $\mathbf{f}(\alpha x)\neq \alpha \mathbf{f}(x)$, input states must have the same normalization as the states in the training set for a meaningful interpretation of the network. To illustrate this statement, we use states from $\mathcal{A}$ multiplied by a factor, i.e., we use $\alpha |\psi|^2$ instead of $|\psi|^2$. Figure~\ref{fig:figure_10} shows the accuracy\footnote{Here we still can talk about the accuracy, since the states $\alpha \psi^2$ correspond to integrable or non-integrable situations.} of the network as a function of $\alpha$. The maximum accuracy of the network is reached at $\alpha=1$, i.e., for the states used for the training. Integrable states are classified as non-integrable almost everywhere except close to $\alpha=1$. A different situation happens for non-integrable states. They are classified correctly almost everywhere, and we conclude that they are less susceptible to the factor $\alpha$. The shape of the curves in Fig.~\ref{fig:figure_10} is not universal, it depends on hyperparameters of the network. However, a general conclusion holds -- the normalization is important, and we use normalized input functions in the further analysis. In principle, it is possible to reduce the sensitivity of our neural network to $\alpha$. To that end, one needs to add states $\alpha|\psi|^2$ to a training set (see data augmentation in deep learning~\cite{shorten2019}). We do not pursue this possibility here, to demonstrate that the sensitivity of integrable states on $\alpha$ is different from that for non-integrable states.

\begin{figure}[t]
\includegraphics[scale=0.6]{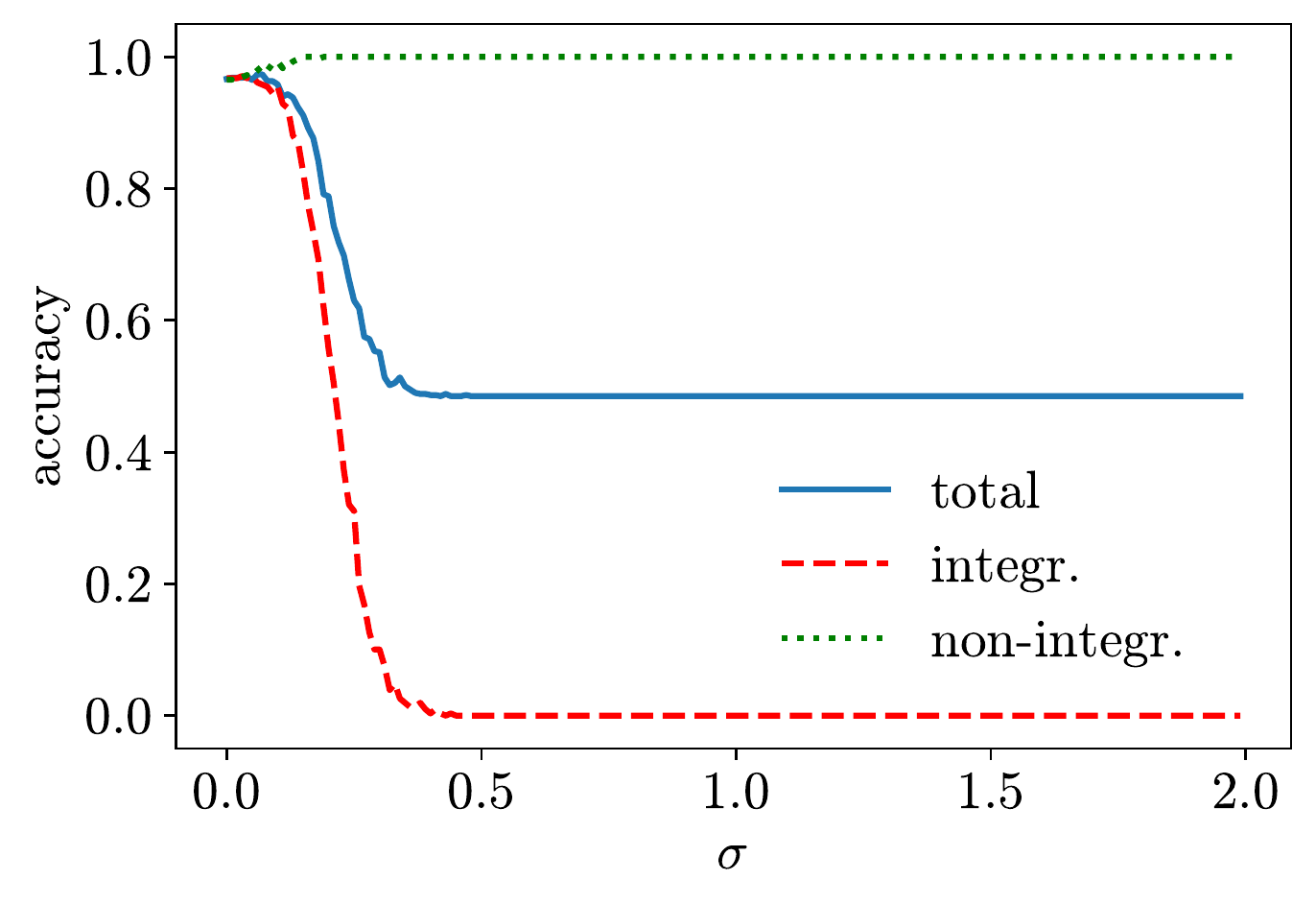}
\caption{Predictions of the network for the states from $\mathcal{A}$ with noise. The (red) dashed curve shows the accuracy for states which are integrable for $\sigma=0$ (the accuracy here is defined as the percentage of the states identified as integrable). The (green) dotted curve shows the accuracy of the network for non-integrable states. The (blue) solid curve shows the average of the first two curves.   
}
\label{fig:figure_11}
\end{figure}

As a next step, we add noise to the images from $\mathcal{A}$, i.e., we build a new data set using wave functions $\tilde \psi=a_\sigma\psi(1+r_\sigma)$, where $r_\sigma$ is a noise function whose values are drawn from the normal distribution with zero mean and the standard deviation $\sigma$; $a_\sigma$ is a normalization factor, which is determined for each input state depending on the function $r_\sigma$. We assume that $r_\sigma$ possesses basic symmetries of the problem: fermionic and mirror. Functions $\tilde \psi$ naturally appear in applications, and therefore, it is interesting to investigate the resilience of the network to random noise. We use 4000 states of $\mathcal{A}$ with noise to make a relevant statistical statement, see Fig.~\ref{fig:figure_11}. Small values of $\sigma$ lead to weak noise and the network correctly classifies almost all input states. 
However, larger values of $\sigma$ lead to confusing input states, and the network fails. It actually fails for integrable states where the noise destroys correlations. The accuracy for non-integrable states is always high. The resilience of the network to noise suggests it as a tool to analyze experimental data (e.g., obtained using microwave billiards). These experiments~\cite{Richter1999} can produce a large amount of data, however, there are limited variety of tools to analyze the simulated states. In particular, neural networks can be used to identify atypical states, which do not fit the overall pattern, e.g., scars.

Our choice of $\tilde \psi$ to represent a noisy state is not unique. One could, for example, use instead of $\tilde \psi$ the function $\bar \psi = a_\sigma(\psi+ \mathcal{G} r_\sigma)$, where the parameter $\mathcal{G}$ determines the relative weight of the function $r_\sigma$, which is defined as above. Note that $\mathcal{G}$ cannot be absorbed in $r_\sigma$, since for a given value of $\sigma$, the average amplitude of $r_\sigma^2$ is well defined.  For $\mathcal{G}=0$, the function $\bar  \psi$ is a physical state, whereas for $\mathcal{G}\to\infty$, the function $\bar  \psi$ is completely random. In contrast to $\tilde \psi$, the function $\bar \psi$ can become completely independent of the function $\psi$. This happens if the parameter $\mathcal{G}$ is large.

\begin{figure}
\includegraphics[scale=0.52]{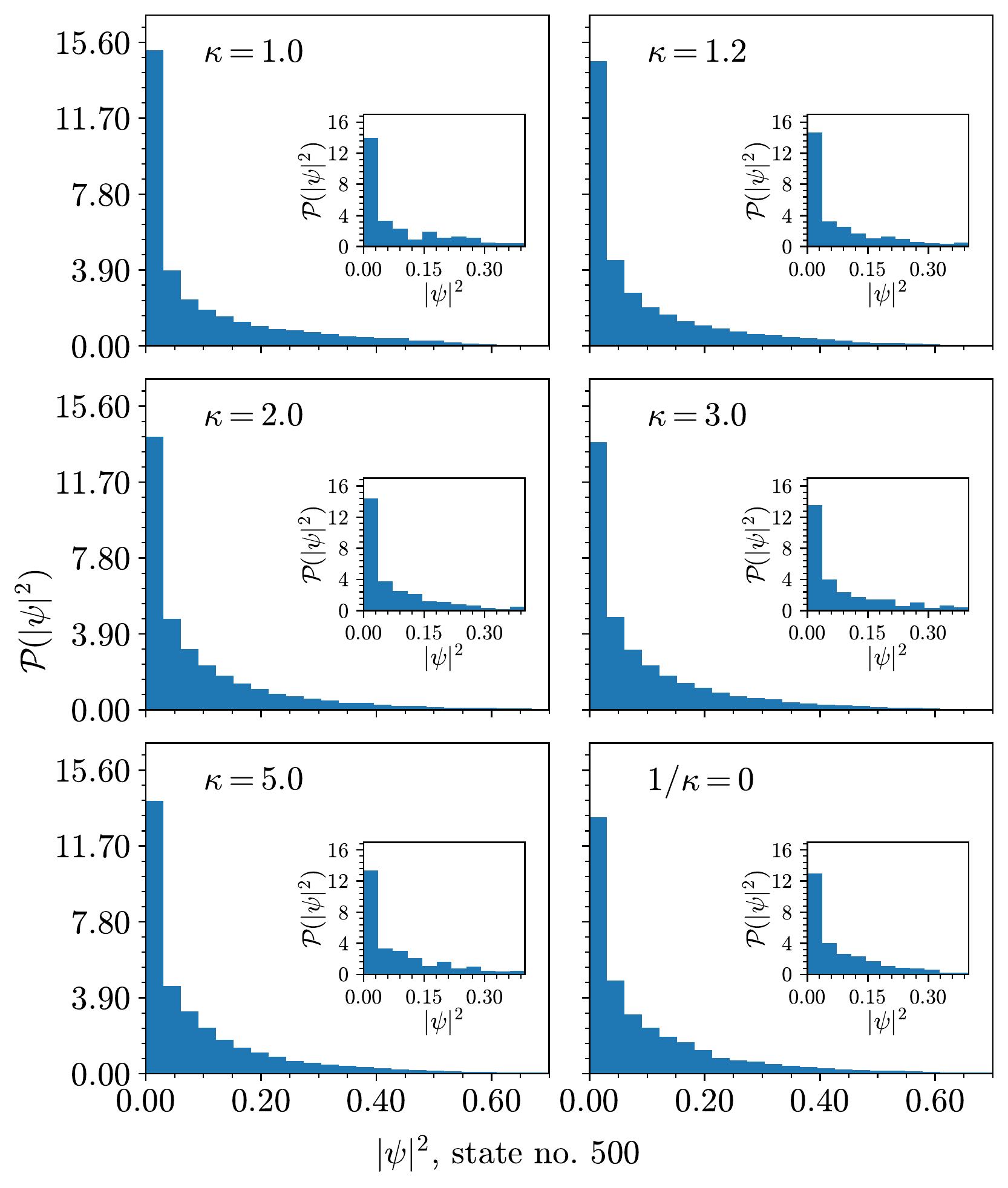}
\caption{The histogram shows distributions of probability-density amplitudes for different values of the mass ratio, $\kappa$. We use a $315\times 315$
 pixel representation of the probability density, $|\psi|^2$, for this analysis. The insets give $\mathcal{P}(|\psi|^2)$  for the probability density at the lower ($33\times 33$) pixel resolution.}
\label{fig:figure_app_1}
\end{figure}

For the data set based upon $\bar \psi$ with small values of $\mathcal{G}$, the accuracy of the network is similar to that presented in Fig.~\ref{fig:figure_11}. For large values of $\mathcal{G}$, the network is confused and classifies states in a random manner. This behavior should be compared with the data set $\tilde \psi$ for which the states are classified as non-integrable when the noise is large. To understand this difference, note that $\tilde \psi$ retains information about the nodal lines of the physical state. It turns out that it is important for the input state to have enough pixels with small (zero) values (note that the number of zero pixels is very large for $|\psi|^2$, see Fig.~\ref{fig:figure_app_1}). Only such states have a direct meaning for the network, all other states confuse the network and do not allow for extraction of any meaningful information. It is worthwhile noting that the network does not learn the physical nodal lines. We checked that almost any random state with a large number ($\simeq 30-40\%$) of zero pixels is classified as non-integrable.

Our discussion above suggests that the network classifies all states with nodal lines but without clear spatial correlations as non-integrable. In other words, the network perceives almost all states as non-integrable, and there are only a few islands recognized as integrable states. Such asymmetry of learning is not expected to happen for labels with equal significance (cat and dog for example).
We speculate that the observed asymmetry is related to the fact that non-integrable states that correspond to different energies do not correlate in space~\cite{Prigodin1995}, and, therefore, a neural network cannot learn any significant patterns when trained with those states. 

All in all, the state classificator validates our expectation that non-integrable states are random and abundant, whereas integrable states are rather unique and occur rarely. 
For example, in our analysis, integrable states are a set of measure zero in $\kappa$, and perturbation in $\kappa$ or noise change the prediction of $\mathcal{N}$ for these states.
The asymmetry also suggests that the  standard implementation of deep learning (DL) presented here should be modified to reveal the physics behind non-integrable states. The present network does not distinguish a non-physical random state with a large number of zero values from a physical non-integrable state since it was not trained for that question. A possible modification is the addition of an extra label ($b_3$ in Eq.~(\ref{eq:mapping_network})) for non-physical states. Since there are many possible non-physical states, one should frame the problem having an experimental or numerical set-up in mind, where such states have some origin and interpretation. We leave an investigation of this possibility for future studies. 

Finally, we note that the conclusion that the network classifies almost all random images as non-integrable is general -- it does not depend on the values of hyperparameters, initial seed or the distribution  (Laplace, Gaussian, etc.) that we use for generation of the random states. We check this by performing a number of numerical experiments. In particular, this means that the network does not learn $\mathcal{P}(\psi)$ from Eq.~(\ref{eq:p_psi}).

To summarize:  A trained network can accurately classify integrable and non-integrable states.  The network can even classify input states from an orthogonal (bosonic) Hilbert space, on which it can have no microscopic information. Thus network identifies generic features of the ‘integrable’
  and ‘non-integrable’ states, although this might be not that surprising provided that we work with low-resolution $64\times 64$ pixel images. The network classifies almost all random images with nodal lines as non-integrable, which suggests that useful information for the network is mostly contained in integrable states.

\section{Summary and Outlook}
\label{sec:concl}

We used convolutional neural networks to analyze states of a quantum triangular billiard illustrated in Fig.~\ref{fig:figure_1}. We argued that neural networks can correctly classify integrable and non-integrable states. The important  features of the states for the network are the normalization of the wave functions and a large number of zero elements, corresponding to the existence of a nodal line. Almost any random image that satisfies these criteria is classified as non-integrable. All in all, the neural network supports our expectation that non-integrable states are resilient to noise as discussed in subsection~\ref{sec:unphys}, and have a `random' structure, unlike integrable states whose structure can be revealed by considering, for example, nodal lines.

Our results suggest that machine learning tools can be used to analyze the morphology of wave functions of highly excited states obtained numerically or experimentally, to solve problems like: find exceptional states (e.g., scars or integrable states) in the spectra, investigate the transition from chaotic to integrable dynamics, etc. However, further investigations are needed to set the limits of applicability for deep learning tools.  For example, our network considers all states without clear correlations as non-integrable. This means that it must be modified for the analysis of noisy data, where a noisy image without any physical meaning could be classified as non-integrable.  To circumvent this, one could introduce additional labels for training the network. For example, one could consider three classes -- `integrable', `non-integrable', and `noise'. This classification might allow for a more precise description of data sets, and may help one to extract more information about the physics behind the problem.

We speculate that a network memorizes integrable states, and all other states are classified as non-integrable, provided that an image has a large number of vanishing values. This would explain our findings and it would align nicely with the observation that we observe no overfitting even after many epochs of working through the training set. In the future, it will be interesting to use other integrable systems to test this idea. In particular, one could use non-triangular billiards, or systems without an impenetrable boundary. For example, one can consider a two-dimensional harmonic oscillator with cold atoms. At a single body level, the integrability in this system can be broken by potential bumps~\cite{Harshman2017a,Heller2019} or spin-orbit coupling~\cite{Marchukov_2014}. 

In the present work, we focus on data related to the spatial representation of quantum states. However, our approach can also be used to analyze other data. For example, for few-body cold atom systems, correlation functions in momentum space are of interest in theory and experiment (see, e.g.,~\cite{Bergschneider2019}).
Therefore, in the future, it will be interesting to train neural networks using experimental/numerical data that correspond to a momentum-space representation of quantum states, and study the corresponding features of `quantum chaos'. Although, in the present work we focused on two-dimensional data (images of probability densities), it might be interesting to consider higher-dimensional analogues (e.g., four-body systems) as well.
Further applications include the identification of phase transitions, the representation of quantum many-body states, and the solution of many-body problems~\cite{DasSarma:2019}. Thus machine learning techniques may not only be used to classify quantum states but also to obtain them by solving the corresponding many-body problem.

To extract further information about the map $\mathbf{f}$, one could investigate its geometry close to its maximum (minimum) values. For example, in the vicinity of some accurately determined integrable state ($f_1(x_0)=1$), we can write
\begin{equation}
f_1(x_0+\delta x)\simeq 1+\frac{1}{2}\delta x^T G \delta x,
\end{equation}
where the position of the maximum $x_0$ is understood as a vector and $G$ is the Hessian matrix. The first derivative of $f_1$ vanishes since the function $f_1$ is analytic and bounded. Eigenstates of the Hessian $G$ provide us with the most important correlations. A preliminary study shows that there are only a handful of eigenvalues of $G$ for our network, which suggests the next step in the analysis of our image classifier.

 A deeper understanding of the way a neural network works
  may be obtained by combining different techniques to interpret its
  operation.  The potential of this approach was illustrated in
  Ref.~\cite{olah2018the}, for instance, by analyzing how a network looking at
  images of a labrador retriever detects floppy ears and how that
  influences its classification.\\

\begin{acknowledgments}
We thank Aidan Tracy for his input during the initial stages of this project. We thank 
Nathan Harshman, Achim Richter, Wojciech Rzadkowski, and Dane Hudson Smith for helpful discussions and comments on the manuscript. 
This work has been supported by European Union's Horizon 2020 research and innovation programme under the Marie Sk\l{}odowska-Curie Grant Agreement No. 754411 (A.G.V.);
by the German Aeronautics and Space Administration (DLR) through Grant No. 50 WM 1957 (O.V.M.); 
by the Deutsche Forschungsgemeinschaft through Project VO 2437/1-1 (Projektnummer 413495248) (A.G.V. and H.W.H.); 
by the Deutsche Forschungsgemeinschaft through Collaborative Research Center SFB 1245 (Projektnummer 279384907) and by the Bundesministerium f\"ur Bildung und Forschung under
contract  05P18RDFN1 (H.W.H).
H.W.H. also thanks the ECT* for hospitality during the workshop
”Universal physics in Many-Body Quantum Systems – From Atoms to Quarks”.
This infrastructure is part of a project that has received funding from the
European Union’s Horizon
2020 research and innovation programme under grant agreement No 824093.
\end{acknowledgments}

\begin{widetext}

\begin{appendix}

\section*{Appendix}
\label{Appendix:1}
To generate the input for a neural network, we diagonalize the Hamiltonian $H_{P=0}$ from Eq.~(\ref{eq:h_p}) in a truncated Hilbert space whose basis element is 
\begin{align}
\xi_{n_1,n_2}(x_1,x_2)=N\left[\sin (\frac{n_1\pi x_1}{L}) \sin(\frac{n_2\pi x_2}{L}) - \sin (\frac{n_1\pi x_2}{L}) \sin(\frac{n_2\pi x_1}{L})\right]\, ,
\end{align}
where $c>n_1>n_2>0$, $c$ is the cutoff parameter, and $N$ is the normalization constant.
In this basis, matrix elements of the Hamiltonian read as (in our numerical analysis, we use units in which $L=\pi$)
\begin{align}
\begin{split}
\int_0^\pi \int_0^\pi  \xi_{m_1,m_2}(x_1,x_2)^* H_{0} \xi_{n_1,n_2}(x_1,x_2) \mathrm{d}x_1 \mathrm{d}x_2&=\frac{1}{2} (n_1^2 + n_2^2)\left( 1+\frac{1}{\kappa}\right)\left( \delta_{m_1,n_1}\delta_{m_2,n_2} -\delta_{m_1,n_2}\delta_{m_2,n_1}\right) \\
& +\frac{n_1 n_2}{\pi^2 \kappa}\left[ I(m_1+n_2,m_2+n_1) + I(m_1+n_2,m_2-n_1)\right.\\
& +I(m_1-n_2,m_2+n_1) + I(m_1-n_2,m_2-n_1)\\
& -I(m_1+n_1,m_2-n_2) - I(m_1+n_1,m_2-n_2)\\
& \left. -I(m_1-n_1,m_2+n_2) - I(m_1-n_1,m_2-n_2) \right]\, ,
\end{split}
\end{align}
where $I(s,t)=\frac{1}{st}\left[(-1)^{s} -1\right] \left[(-1)^{t} -1\right]$ if $s,t\neq 0$, and $I(s,t)=0$ otherwise.
To write these matrix elements in a matrix form, we use the index
\begin{align}
n=n_2-n_1 +c(n_1-1) -\frac{(n_1 -1)n_1}{2} \, .
\end{align}
The parameters $n_1$ and $n_2$ can be uniquely defined as:
\begin{align}
\begin{split}
n_1 = \frac{1+2c}{2} - \sqrt{c^2 -c -2n +\frac{9}{4}}, \qquad
n_2 =n+n_1 -c(n_1 -1) +\frac{(n_1-1)n_1}{2} \, .
\end{split}
\end{align}
To choose the cutoff parameter, $c$, we should find a good balance between the calculation time and the accuracy of our results. To quantify the accuracy, we compute energies for $\kappa=1$ by diagonalizing $H_{0}$, and compare them to the exact ones obtained with the Bethe ansatz:
\begin{align}
E_{BA}=
\begin{cases}
n_1^2 +n_2^2 + n_3^2, & \sum n_i=0,\\
(n_1+\frac{1}{3})^2 +(n_2+\frac{1}{3})^2 + (n_3+\frac{1}{3})^2, & \sum n_i=-1,\\
(n_1+\frac{2}{3})^2 +(n_2+\frac{2}{3})^2 + (n_3+\frac{2}{3})^2, & \sum n_i=-2.
\end{cases}
\end{align}
This solution assumes that the total momentum is zero.
The relative difference
\begin{align}
\epsilon=\frac{E_{BA}-E(c)}{E_{BA}}
\end{align}
provides us a measure for the accuracy. Note that our exact diagonalization method is expected to work better for $\kappa>1$, since $\xi$ is the eigenstate of a system with $1/\kappa=0$. The input for a neural network is obtained using $c=130$, for which we obtain 726 (3795) states with $\epsilon<10^{-4}~(10^{-3})$ within a short enough computation time.

\end{appendix}

\end{widetext}

\bibliography{bib} 

\end{document}